\documentclass[a4paper,UKenglish,cleveref, autoref]{lipics-v2019}

\newcommand{\norm}[1]{\left\lVert#1\right\rVert}

\usepackage{xpunctuate}
\newcommand*{\eg}{e.g\xperiod}
\newcommand*{\ie}{i.e\xperiod}
\newcommand*{\etal}{\textit{et al}\xperiod}

\title{Quantified uncertainty of flexible protein-protein docking algorithm}
\titlerunning{Uncertainty quantified protein docking}

\author{Nathan Clement\footnote{nclement@cs.utexas.edu}}{Department of Computer Science, University of Texas at Austin, USA}{}{}{}

\authorrunning{N.\,L. Clement, A. Gomes, and C.\,L. Bajaj}

\Copyright{Nathan L. Clement, Abel Gomes, and Chandrajit L. Bajaj}



\category{}

\acknowledgements{I would like to thank all those who have supported and helped advise on this work, for their valuable feedback and suggestions for improvement.}

\nolinenumbers 

\hideLIPIcs  

\EventEditors{John Q. Open and Joan R. Access}
\EventNoEds{2}
\EventLongTitle{42nd Conference on Very Important Topics (CVIT 2016)}
\EventShortTitle{CVIT 2016}
\EventAcronym{CVIT}
\EventYear{2016}
\EventDate{December 24--27, 2016}
\EventLocation{Little Whinging, United Kingdom}
\EventLogo{}
\SeriesVolume{42}
\ArticleNo{23}

\begin{document}

\maketitle

\begin{abstract}
The strength or weakness of an algorithm is ultimately governed by the confidence of its result. When the domain of the problem is large (\eg traversal of a high-dimensional space), a perfect solution cannot be obtained, so approximations must be made. These approximations often lead to a reported quantity of interest (QOI) which varies between runs, decreasing the confidence of any single run. When the algorithm further computes this final QOI based on uncertain or noisy data, the variability (or lack of confidence) of the final QOI increases. Unbounded, these two sources of uncertainty (algorithmic approximations and uncertainty in input data) can result in a reported statistic that has low correlation with ground truth.

In biological applications, this is especially applicable, as the search space is generally approximated at least to some degree (\eg a high percentage of protein structures are invalid or energetically unfavorable) and the explicit conversion from continuous to discrete space for protein representation implies some uncertainty in the input data. This research applies {\it uncertainty quantification} techniques to the difficult protein-protein docking problem, first showing the variability that exists in existing software, and then providing a method for computing {\it probabilistic certificates} in the form of Chernoff-like bounds. Finally, this paper leverages these probabilistic certificates to accurately bound the uncertainty in docking from two docking algorithms, providing a QOI that is both robust and statistically meaningful.
\end{abstract}

\section{Introduction}
Predicting the correct binding formation of two proteins (protein-protein docking) has many applications in medicine and biology \cite{Rasheed2015gp120,lapointe2013structural}. The simpler form of this problem is the so-called ``bound-bound case,'' where the 3-dimensional coordinates of the {\it in situ} protein complex is resolved (via \eg X-ray crystallography, NMR, etc.), and atoms corresponding to individual proteins are then extracted from the complex. The more difficult version is the unbound-unbound case, where each protein in the pair is imaged in its separate native state, and the algorithm must predict the correct {\it in situ} bound complex \cite{kuroda2016pushing}. Importantly, the final quantity of interest (QOI) in many cases is the change in bound free energy: protein complexes with a high change in free energy are more likely to be found as a bound complex, and are likely good targets for drug discovery pathways. The difficulty of the unbound-unbound case then arises from the inherent flexibility of proteins: large-scale movements may occur along the pathway from a closed conformation (unbound) to an open (bound) one, or visa versa. If docking is performed on only the unbound complexes, the final delta energy could be completely false.

This paper provides a framework for discussing the difficulties in computing the final QOI when starting with the unbound conformation of pair of proteins. Specifically, we generate a low-discrepancy sample of potential proteins, and repeatedly dock each of these proteins with a given piece of software. Once we have performed this several times, we report the final QOI as a certificate. We show that this certificate includes the true QOI in most cases. To aid in discussion, here and through the paper we will refer to one protein (typically the larger) as the {\it receptor}, and the other as the {\it ligand}.

One of the major difficulties of the unbound-unbound docking problem is that computational approaches must search two high-dimensional spaces. The first is that of possible protein structures, a naive description of which is $\mathbb{R}^{3*n_L}\times\mathbb{R}^{3*n_R}$, where $n_L$ and $n_R$ are the number of atoms in the ligand and receptor, respectively. The second is the space of possible docked conformations. In rigid-body docking (each protein is static), this is the 6-dimensional real space of 3 rotational + 3 translational degrees of freedom, $SE(3)=SO(3)\times \mathbb{R}^3$ \cite{f2dock,padhorny2016protein}. Without any approximations, searching this high-dimensional space is computationally intractable. To achieve meaningful results, successful algorithms must employ some sort of approximations.

The approach using the least amount of approximations in either space is {\it molecular dynamics} (MD) simulations, such as AMBER \cite{amber2016}. These approaches use Neutonian equations of motion to minimize an energy function based on physical properties of the system (either in solute or alone). However, these simulations suffer from numerical approximations, and are usually only valid for very short timescales, representing only short-ranged motions \cite{klepeis2009long}. This is not often the case with the movements of unbound-unbound protein docking, which can vary widely over a long period of time (in the range of ms to seconds \cite{lapointe2013structural}).

Instead, simplifications to both of these spaces can be made, either separately or together. The most extreme example of this is rigid-body docking, where the assumption is made that the first space (protein conformational space) is empty. For proteins which do not undergo a large amount of motion while docking, this is a good approximation and can lead to meaningful results \cite{f2dock,F2DockGBRerank}. For more difficult protein-protein docking examples, most programs use approximations in both spaces, using a simplified model of the protein as well as a non-exhaustive representation of the docked search space, either by coarsening or randomization approaches.

One of the biggest issues that arises from these simplifications is {\it uncertainty propagation}. A computational representation of a protein is, by nature, an approximation (discrete representation of a continuous space). Computing a simple statistic, or {\it quantity of interest} (QOI), on these representations is then by nature uncertain \cite{rasheed2017uq}. Algorithmic approximations (due to randomness or variations in the inputs) in one stage of a protein docking pipeline lead to uncertainty in the input for the next stage. If these uncertainties are not {\it quantified} at each stage, the uncertainties propagate to future levels of the pipeline, leading to a result or QOI that is unbounded, and may contain little valuable information.

This research provides a framework for bounding the uncertainty of protein-protein docking. For a docking procedure where the QOI, $f(\mathbf{X})$, is some complicated function or optimization functional involving noisy data $\mathbf{X}$, we seek to provide a {\it probabilistic certificate} as a function of parameter $t$ that the computed value $f(\mathbf{X})$ is not more that $t$ away from the true value, with high probability. This certificate is expressed as a Chernoff-Hoeffding like bound \cite{Chernoff_1952} as follows:
\begin{equation}
  \Pr\left[\left|f(X)-E[f]\right|>t\right]\le\epsilon,\label{eqn:chernoff}
\end{equation}
where $E[f]$ is the expectation of $f$, computed over all permutations of $X$.

The primary QOI we are interested in bounding is the change in Gibbs free energy, or $\Delta G$, as this is the metric most useful for real-world experiments. However, we also consider the interface RMSD (iRMSD), which is defined as the RMSD between $C\alpha$ atoms on the interface of the bound pair. We consider proteins individually, and use the proteins labeled ``Diffilult'' from the ZLab benchmark 5 \cite{zlab5}.

Instead of providing a new docking algorithm as a solution to bounding the above certificate, this research instead considers the docking algorithm, $f(\mathbf{X})$, as a {\it black box}, exploring the landscape of possible structures, $X\in\mathbf{X}$, as inputs to $f$ and computing the certificates from the output. This then provides a framework by which any two algorithms can be compared, and by which conclusive results can be reported.

%
%

First, we show that the protein sampling protocol used improves upon the results of both rigid-body and flexible docking metrics, then, we compute probabilistic certificates for the change in Gibbs free energy for sets of docked proteins. Finally, we discuss the importance of these results both in terms of UQ for docking algorithms and in biological relevance.

\subsection{Related work}
The major guiding factor in the exploration of protein space is that, while the naive degrees of freedom are large, the number of degrees of freedom necessary to represent the majority of protein movement can be described using only a small number of constraints, for example the principle eigenvector \cite{olson2012guiding}, a 10-dimensional projection \cite{novinskaya2015enmin}, or only the first few (1-3) modes from a normal modes analysis \cite{hinsen2000harmonicity,tama2000building,dobbins2008insights}.

A more difficult problem is {\it de novo} protein assembly from fragments. The goal is to explore the conformational landscape of the protein, with a focus on identifying the energy minima. Approaches to this include singular value decomposition \cite{mustard2005docking,olson2012guiding}, normal modes analysis (with distance constraints \cite{zheng2006modeling}), projections onto a lower dimensional space (guided by an ``expert'', e.g. human-generated, projection \cite{novinskaya2015enmin}), and probabilistic graphical models \cite{Zhao2008discriminative,Zhao2010fragment,bhattacharya2015novo}. The sample generation employed in this work is somewhat different in two aspects: 1) instead of starting {\it de novo}, we start with a given protein structure (the unbound configuration), and 2) instead of trying to generate the low-energy minima, we recognize that individually, the proteins in the bound conformation are likely at a higher energy state that their unbound counterpart, so we generate both good and bad samples in the local configurational space.

With some small amounts of information (a small number of distance \cite{zheng2006modeling} or compactness \cite{Hamelryck2006} constraints, or a predicted value of the ``radius of gyration'' \cite{uyar2014features}), a protein close to the bound conformation can usually be found, and is even possible for specific hinge proteins with no additional information \cite{yesylevskyy2008blind}. Most of these representations use a graph representation of the protein with one node per atom; however, other research has shown that the normal modes from a NMA model with all atoms is about the same as $C\alpha$ only \cite{schuyler2004normal}, or even coarser using an RTB approach \cite{hoffmann2017nolb,tama2000building}.

Amadei~\etal \cite{amadei1995essential} suggested it was possible to separate the protein configurational space into two subspaces, the ``essential'' subspace (only a few degrees of freedom that give rise to the large-scale motions of the protein) and the ``physically constrained'' subspace, consisting of largely Gaussian motion. Thus, identifying a small number of ``hinge'' residues that give rise to the 'essential' subspace has been the focus of another group of work. Shamsuddin~\etal \cite{Shamsuddin2014} constructed $d$ basis vectors to define the motion of rigid bodies joined by hinge residues, HingeProt\cite{emekli2008hingeprot} is a popular method for identifying hinges using cross-correlation of movements from a normal modes analysis (NMA).

The observation that the space of $(\psi, \phi)$ protein torsion angles is dependent on the secondary structure and amino acid type was first made by Ramachandran in 1963 \cite{Ramachandran1963}. Since this time, other research has expanded on this idea and applied it to protein structure generation. Ting~\etal created a library of Ramachandran distributions of loop torsion angles, generated from a high-quality set of proteins and represented with a hierarchical Dirichlet process model \cite{Ting2010}. Other libraries for protein distributions also exist, such as the energy-based side-chain library from Subramaniam and Senes \cite{subramaniam2012energy}. In this work, we generate our own set of neighbor-dependent Ramachandran distributions, but also convolve them with a bivariate von Mises distribution centered at the torsion angles from the original protein. This allows us to generate protein samples that are more consistent to the input protein, but still allow for natural flexibility.

\subsubsection{Docking protocols}
There are a wide variety of docking protocols that are used for docking proteins. In this paper, we will compare three programs:
\begin{itemize}
  \item F2Dock\cite{f2dock}, a rigid-body docking algorithm. F2Dock uses convolution in the Fourier space to reduce the complexity of the space of possible docked conformations. This initial docking returns a set of poses that have high structural overlap. After this initial stage is done, F2Dock uses statistical potentials that approximate the Generalized Born (GB) energy model to rerank each initial pose \cite{F2DockGBRerank}. F2Dock is a rigid-body docking algorithm, and does not consider fluctuations in protein conformations.
    \item RosettaDock\cite{rosettadock,rosettadock_benchmark} is a semi-flexible docking algorithm which employs two levels of protein representation to find an optimal docked conformation. At the first stage, RosettaDock uses rigid-body permutations of a coarse representation of the protein (side-chain atoms as a single pseudo-atom), followed by a high-resolution second stage with all-atom representation of the protein allowing for side-chain flexibility with side-chain packing and energy minimization.
    \item SwarmDock\cite{swarmdock_first} is a flexible docking algorithm, and represents proteins internally with a RTB (rotation-translation of blocks) methodology, for both the receptor and the ligand. The RTB provides a natural method for decomposing the protein into ``building blocks,'' which are the basis for nomal modes analysis (NMA). The docking algorithm is a {\it mimetic algorithm}: a genetic algorithm coupled with a learning procedure to perform local refinements.
\end{itemize}

\subsubsection{Uncertainty quantification}
In previous research \cite{rasheed2017uq}, we have shown how individual QOI can vary depending on uncertainty in the input structure, namely atomistic positional variations arising from X-ray crystallography and represented as B-factors. We treated each atom as a random variable, and sampled from the product space of $N$ independent gaussians, where $N$ is the number of atoms in the protein. We showed that, under this simplified model of small perturbations, there can be a wide variance in simple QOI. In \cite{clement2018viral}, we further showed that perturbations under this simplistic model can propagate to further uncertainties in the viral assembly problem.

In this work, we expand upon our previous research in the following manner. First, the model used in the previous research was simplistic, and, while useful for modeling small uncertainties, does not provide insight into uncertainty of large-scale protein movement. Second, we consider the impact of this conformational uncertainty to provide certificates for black-box docking functions. This second contribution can be used when trying to interpolate results of a given docking algorithm to biological equivalents.

The only known research that applies uncertainty quantification to protein-protein docking is a recent preprint by Cau and Shen \cite{cao2019bayesian_preprint}. The authors use Bayesian active learning to explore protein-protein docking samples using a black-box energy function. Once the energy landscape has been sufficiently sampled, they provide posterior distributions of the desired QOI, which enables computing confidence intervals for each model. The protein sampling procedure uses a combination of traditional NMA and ``complex'' normal modes \cite{oliwa2015cnma}, and samples are guided in the direction of each of these modes. The major differences between this work and our own work is 1) the treatment of the entire {\it docking algorithm} as a black box (instead of just the energy function, and 2) the use of a hierarchical model convolved with a von Mises distribution to generate samples local to the unbound input.

\section{Materials and methods}
In this section, we provide the theoretical and technical details of our approach. First, we will provide the theoretical basis of our work and show how we can provide probablistic certificates through effective sampling. Next, we will describe our protein representation and sampling protocol. Finally, we will describe the benchmark dataset we use in this work.

\subsection{Computing Chernoff-like bounds}
Our primary motivation in this work is to compute a {\it probabilistic certificate} to bound the uncertainty in a computed statistics. We are most concerned with providing the Chernoff-Hoeffding like bound expressed in Equation~\ref{eqn:chernoff}, which provides a probabilistic guarantee for the moments of a QOI computed on noisy data.

We can provide a theoretic bound for the uncertainty by using the McDiarmid inequality, defined in \cite{McDiarmid_1989} and extended to support summations of decaying kernels such as the Leonnard-Jones potential in \cite{rasheed2017uq}. Let $(X_i)$ be independent random variables with discrete space $A_i$, let $f:\Pi_iA_i\rightarrow \mathbb{R}$, and let $|f(x_1,\ldots,x_k,\ldots,x_n)-f(x_1,\ldots,x_k',\ldots,x_n)|\le c_k$, or $c_k$ is the degree of change influenced on $f$ over all variations of $x_k$. Then, for any $t > 0$:
\begin{equation*}
  \Pr\left[\left|f(\mathbf{X})-\mathbb{E}\left[f\right] \right|>t \right] < 2\exp\left(-2t^2/\sum_kc^2_k\right).
\end{equation*}

Thus, to provide theoretic bounds, all that is required is to determine the value of $c_k$ for each $x_k$. However, computing $c_k$ analytically may be difficult, and even if it were possible, theoretical bounds these often overestimate the error. An alternate approach is then to empirically compute these certificates using quasi-Monte Carlo (QMC) methods \cite{Niederreiter_1990,James_Hoogland_Kleiss_1996}. Assuming the distribution of $(X_i)$ is known, we sample this space and evaluate $f$ at each sample. This leads to an estimate of the distribution of $f$ over the joint space of all $A_i$, which provides sufficient data to compute certificates on the uncertainty, as is defined in Equation~\ref{eqn:chernoff}.

Correctness of this approach relies on the correctness of the QMC methods and the description of the joint sampling space. For this reason, we will spend the next section describing our protein representation and the corresponding sampling space. In the Results and discussion section we will show that our sampling space is accurate (\eg a good representation of the distribution of $(X_i)$), and thus the certificates we provide are also sound.

\subsection{Protein representations}
\begin{figure}
  \includegraphics[width=0.3\linewidth]{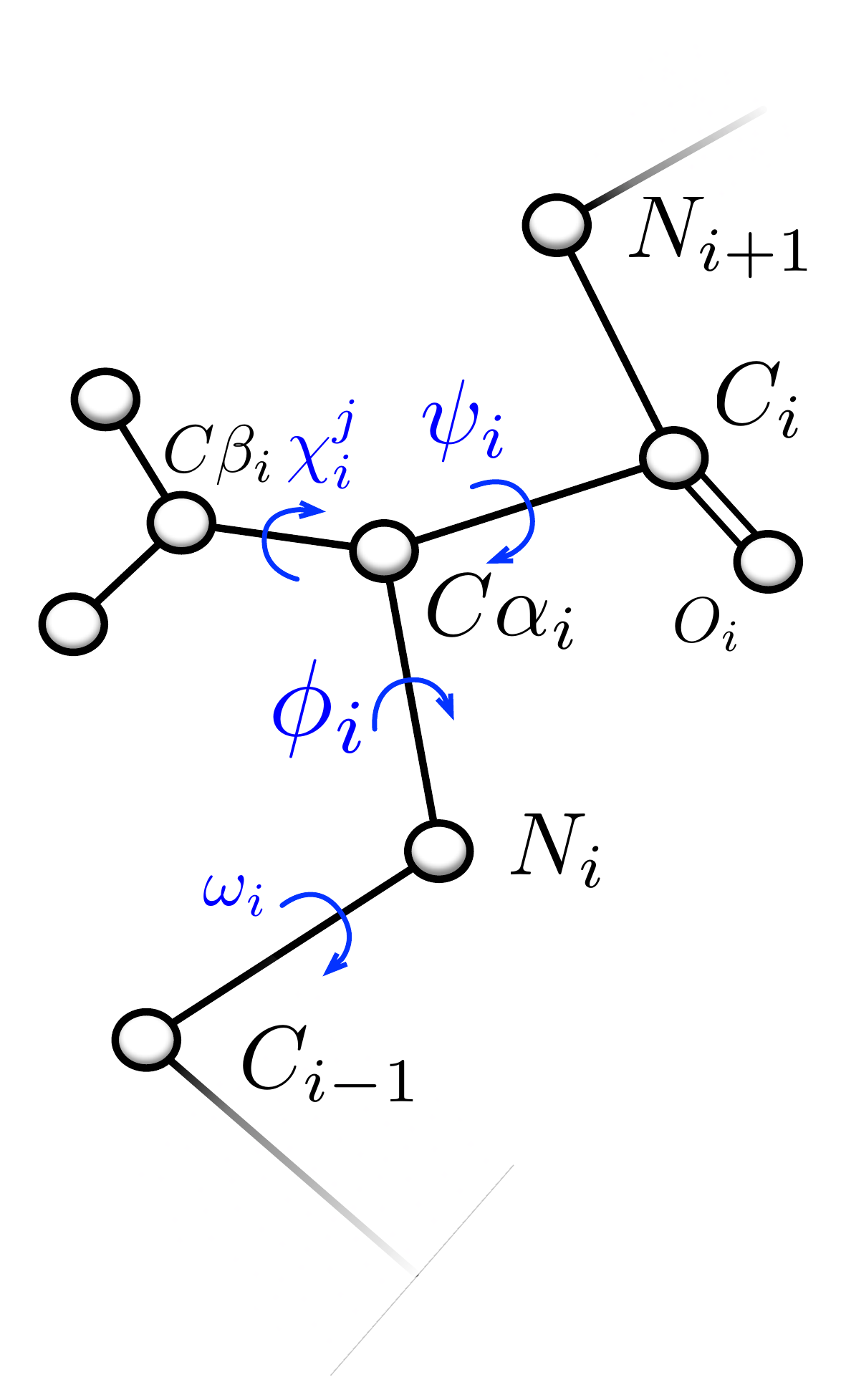}
  \caption{Torsion angles for a protein chain. The backbone atoms are labeled $C_i$, $O_i$, $C\alpha_i$, and $N_i$. For a constrained internal coordinate representation, only the $\psi_i$, $\phi_i$, and potentially $\chi_i$ torsion angles are considered ($\omega_i$ is fixed at 180$^\circ$).}\label{fig:torsions}
\end{figure}

The base structure of a proteins is a linear chain of {\it amino acids} (also called ``residues'') forming a 3-dimensional structure. Each amino acid consists of a set of atoms, and all the atoms connected by covalent bonds into a single 3-dimensional structure. Atoms in amino acids are divided into two groups: {\it backbone} atoms: two carbons, one nitrogen, and one oxygen; and zero or more {\it side-chain} atoms. The carbon connecting the backbone to the side-chain atoms is called the $C\alpha$ atom, and the first side-chain carbon (if it exists) is called the $C\beta$ atom (see Figure~\ref{fig:torsions}). There are 20 different amino acid types, which are defined based on the content and connectivity of their sidechain atoms.

At a high level of description, a single protein can have primary, secondary, and tertiary structure. The sequence of specific amino acids is called the protein {\it primary structure}. The {\it secondary structure} of a protein consist of long-range interactions (non-covalent bonds) between amino acids, typically as helices or sheets. The {\it tertiary structure} of a protein typically defines the biological properties of the protein, and consists of large-scale interactions between secondary structure elements.

The native representation of a protein is a graph in 3-dimensional Cartesian space, where each node of the graph represents atoms and edges represent bonds. The position of each node/atom is represented by a vector in $\mathbb{R}^3$, requiring 3 parameters for each atom. If $\hat t$ is the average number of side-chain atoms per residue for a given protein, then this representation requires a total of $n=3(\hat t+4)N$ parameters, or degrees of freedom (DOFs), for a protein with $N$ amino acids.

\subsubsection{Constrained Internal Coordinate Representation}
An alternate representation of proteins, employed by most sampling protocols (e.g. \cite{Milgram2008, Gangupomu2013} and others) is the {\it internal coordinates} representation. Under this representation, the position of each atom is only defined {\it in relation} to the atoms around it, and the free variables are bond angles, bond lengths, and torsion angles (the degree of ``twist'' defined by 2 planes or 4 atoms, see Figure~\ref{fig:torsions}). This does not immediately reduce the total degrees of freedom (since in general, each {\it atom} needs to be described by bond angles, bond lengths, and torsion angles); however, if small-scale atomic vibrations are ignored, then bond lengths and angles can be approximated as constant, leaving the only DOFs as the $\psi$, $\phi$, and $\chi_i$ torsion angles (the $\omega$ torsion angle on the backbone is held at $\sim 180^\circ$ by the $sp^2$ partial double bond \cite{Brown2009}). If $\hat{k}$ is the average number of $\chi_i$ angles for a given residue ($k$ varies from 0 to 5 in the standard 20 amino acids), then the number of DOFs for this representation for a protein with $N$ amino acids is $m=(\hat{k}+2)N$. Since in most cases $\hat{k}+2\ll 3(\hat{t}+4)$, this {\it constrained} internal coordinate method allows for a lower-dimensional specification of the protein conformational space without a loss in representation \cite{Gipson2012}.

\subsection{Normal modes analysis}
\label{sec:nma}
%
The Gaussian network model (GNM) can be used as a minimal model to describe protein fluctuations. Let $\mathbf{r}_i^0$ represent the {\it equilibrium} position of atom $i$, and let $\mathbf{r}_i$ represent the {\it instantaneous} position of the same atom, where $\mathbf{r}_i=\begin{bmatrix}x_i & y_i & z_i\end{bmatrix}$ (\ie a row vector $\mathbf{r_i}\in \mathbb{R}_3$). Then $\Delta \mathbf{r}_i = \mathbf{r}_i-\mathbf{r}_i^0$ is the fluctuation or deviation vector from the equilibrium position. We will also define $s_{ij}$ as the instantaneous difference between atoms $i$ and $j$ (\ie the spring length), and $\Delta\mathbf{s}_{ij}=\mathbf{s}_{ij}-\mathbf{s}_{ij}^0=\Delta\mathbf{r}_i-\Delta\mathbf{r}_j$ as the vector of deviations in distances, or the vector of relative fluctuations. Then the potential energy of the system of $N$ atoms according to the GNM model, $V_{GNM}$, can be written as \cite{rader2006gaussian}:
\begin{align}
 V_{GNM} & = \frac{\gamma}{2}\left[\sum_{i,j}^N \Gamma_{ij}\norm{\Delta \mathbf{s}_{ij}}^2 \right]\nonumber\\
   & = \frac{\gamma}{2}\left[\sum_{i,j}^N \Gamma_{ij}\left[ (\Delta x_j - \Delta x_i)^2 + (\Delta y_j - \Delta y_i)^2 + (\Delta z_j - \Delta z_i)^2\right] \right],\label{eqn:vgnm_1}
\end{align}
where $\gamma$ is a force constant (uniform for all springs), and $\Gamma_{ij}$ is the $ij$th element of the Kirchoff (connectivity) matrix, defined as:
\begin{equation}
 \Gamma_{ij} =
 \begin{cases}
  -1, & \text{if } i \ne j \text{ and } \norm{\mathbf{s}_{ij}} \le r_c \\
  0, & \text{if } i \ne j \text{ and } \norm{\mathbf{s}_{ij}} > r_c \\
  -\sum_{k\ne i} \Gamma_{ik}, & \text{if } i = j
 \end{cases}\label{eqn:kirchoff}
\end{equation}
where $r_c$ is a cutoff distance between two atoms.

If we then define $\Delta R$ to be the $N\times 3$ matrix of $\Delta \mathbf{r}_i$ values (each $\Delta \mathbf{r}_i$ is a row vector), then we can define $\Delta\mathbf{x}$, $\Delta\mathbf{y}$, and $\Delta\mathbf{z}$ as the $N$-dimensional fluctuation vectors corresponding to each of the $N$ values of $\Delta\mathbf{r}_i$ (\eg $\Delta\mathbf{x}=\Delta R_{:1}$). Using this notation, Equation~\ref{eqn:vgnm_1} becomes:
\begin{equation}
  V_{GNM} = \frac{\gamma}{2}\left[\Delta \mathbf{x}^\intercal \Gamma \Delta \mathbf{x}^\intercal + \Delta \mathbf{y}^\intercal \Gamma \Delta \mathbf{y} + \Delta \mathbf{z}^\intercal \Gamma \Delta \mathbf{z} \right].\label{eqn:vgnm}
\end{equation}

If the residue fluctuations, $\Delta R$, are modeled as a multivariate Gaussian, then the {\it correlation of atomic fluctuations} can be written as \cite{rader2006gaussian}:
\begin{equation}
  \left\langle \Delta r_i \cdot \Delta r_j\right \rangle = \frac{3k_BT}{\gamma}\left[\Gamma^{-1}\right]_{ij}.\label{eqn:cross_gen}
\end{equation}

Further, consider the eigenvalue decomposition of $\Gamma=U\Lambda U^\intercal$, then $U^\intercal=U^{-1}$, the columns of $U$, $\mathbf{u}_i$, are the eigenvectors of $\Gamma$, and the entries of the diagonal matrix $\Lambda$, $\lambda_i$, are the eigenvalues of $\Gamma$. Each eigenvalue corresponds with one mode of movement of the protein. The first (smallest) eigenvalue corresponds to the trivial mode (all atoms moving together), but the following $N-1$ eigenvectors (eigenvalues) correspond to the {\it frequencies} (shape) of individual modes. We can finally rewrite the cross correlation between two atoms $i$ and $j$ as the sum over the $N-1$ nonzero modes using Eq.~\ref{eqn:cross_gen} as:
\begin{equation}
  \left\langle \Delta\mathbf{r_i} \cdot \Delta\mathbf{r_j}\right\rangle = \frac{3k_BT}{\gamma}\sum_{k=2}^N \left[\lambda_k^{-1}\mathbf{u}_k\mathbf{u}_k^\intercal\right]_{ij}
\end{equation}

Additionally, we can use these eigenvectors to describe the {\it cross-correlation between residue fluctuation according to mode $k$} as:
\begin{equation}
 [\Delta \mathbf{r}_i \cdot \Delta \mathbf{r}_j]_k = \frac{3k_BT}{\gamma}\lambda_k^{-1}[\mathbf{u}_k]_i[\mathbf{u}_k]_j\label{eqn:ms_cross},
\end{equation}
and the $N\times N$ mean square fluctuation matrix according to mode $k$, $\left[F\right]_k$, is then:
\begin{equation}
 [F]_k = \frac{3k_BT}{\gamma}\lambda_k^{-1}\mathbf{u}_k\mathbf{u}_k^\intercal.\label{eqn:cc_mat}
\end{equation}
We can also define the {\it vibrational frequency} of a given mode, $k$, $\omega_k$, as:
\begin{equation}
 \omega_k = \left(\gamma\lambda_k\right)^{1/2}.\label{eqn:vib_freq}
\end{equation}

\subsubsection{Anisotropic Network Model (ANM)}
One of the limitations of the GNM is that it can only model relative movements, and not provide the direction of these movements as well. The ANM models each atom as an {\it anisotropic} Gaussian, and includes measures of force in each direction \cite{rader2006gaussian}. Instead of using the Kirchoff matrix from Eq.~\ref{eqn:kirchoff}, a $3N\times M$ cosine matrix $B$ is used:
\begin{align}
  B_{3i+0,j} & = \sum_{j\ne i}f_{ij}\cos\left(\alpha_{ij}^{(X)}\right)h\left(r_c-\norm{\mathbf{s}_{ij}}\right) & = \sum_{j\ne i} f_{ij}(x_j-x_i)/\norm{s_{ij}}h(r_c-\norm{\mathbf{s}_{ij}}) \nonumber\\
  B_{3i+1,j} & = \sum_{j\ne i}f_{ij}\cos\left(\alpha_{ij}^{(Y)}\right)h\left(r_c-\norm{\mathbf{s}_{ij}}\right) & = \sum_{j\ne i} f_{ij}(y_j-y_i)/\norm{s_{ij}}h(r_c-\norm{\mathbf{s}_{ij}}) \nonumber\\
  B_{3i+2,j} & = \sum_{j\ne i}f_{ij}\cos\left(\alpha_{ij}^{(Z)}\right)h\left(r_c-\norm{\mathbf{s}_{ij}}\right) & = \sum_{j\ne i} f_{ij}(z_j-z_i)/\norm{s_{ij}}h(r_c-\norm{\mathbf{s}_{ij}})\label{eqn:cos_B}
\end{align}
where $\alpha_{ij}^{(X)}$ is the angle between the $X$-axis and the vector $\mathbf{s}_{ij}$ (the spring connecting atoms $i$ and $j$), $f_{ij}$ is the force on atom $i$ due to its interaction with atom $j$, and $h\left(r_c-\norm{\mathbf{s}_{ij}}\right)$ is the heavyside step function equal to 1 if the value is 0 (\ie enforcing the summation to only include atoms within the cutoff distance of $r_c$).

The eigenvalue decomposition of $B$, \eg $BB^\intercal=U\Lambda U^\intercal$, then leads to $3N$ eigenvalues, with the first 6 modes being trivial (3 rotational and 3 translational movements that do not change energy). The eigenvectors corresponding to each of the remaining $3N-6$ modes correspond to the $x$- $y$- and $z$-directional forces. Additionally, it can be shown \cite{atilgan2001anisotropy} that $BB^\intercal$ is equivalent to the $3N\times 3N$ potential energy Hessian matrix, which then can extend to more advanced potential energy equations, such as the Amber potential \cite{hinsen2000harmonicity}.


\subsubsection{Practical limitations of NMA}
In practice, there are a few guidelines for applying NMA. First, while the modes from the ANM provide directionality of movement, the smaller modes of the GNM model correspond better with the broader movements of the actual protein \cite{atilgan2001anisotropy}. For this reason, a typical approach is to compute the slowest modes with the GNM, then find the corresponding modes in the ANM with the same fluctuation value, $\omega_i$, to obtain the direction of movement. HingeProt\cite{emekli2008hingeprot} and others (see, for example \cite{atilgan2001anisotropy}) use this approach.

A more important limitation of using NMA arises from assumptions made in the model. It is quite obvious that the atomic fluctuations describe only {\it instantaneous} fluctuations, and since the input model is assumed to be at steady-state, they only describe the instantaneous fluctuations of this configuration. Without additional inputs, structural extrapolation from NMA alone is extremely limited. Second, if the eigenvalue of two modes is equal (or nearly equal), it is quite likely that this similarity is due to numerical error or approximation, and is not due to a signal from the model. Finally, as the frequency (Equation~\ref{eqn:vib_freq}) of the mode increases, the {\it distinguishability} of a given mode decreases. The modes in the higher frequency spectrum ($\omega_i>$85ps$^{-1}$) correspond to the higher-frequency vibrations of hydrogen atoms and the mid-range frequencies correspond to vibrations of single amino acids or secondary structure elements. Only a few low-frequency modes can provide reasonable representations for biologically-relevant movements \cite{hinsen2000harmonicity,tama2000building,dobbins2008insights}.

\subsection{Sampling protocol}
In this section, we will define the hierarchical representation we use to describe a protein, and will also describe how we obtain a set of representative samples of the protein.

\subsubsection{Hierarchical protein representation}
When representing large-scale protein motion, we are primarily interested in ``hinge residues,'' or flexible residues connecting large mostly-rigid bodies. However, since there may be multiple levels of motion, we use a {\it hierarchical representation} of the constrained internal angles representation of the protein. The hierarchical representation is not a recursive subdivision of the protein, but rather a description of specific protein motions. This allows us to represent motions at one level that consist of atoms from different domains in the previous level.

To obtain the hinge residues, we model the protein as a $C\alpha$ (one node per residue) GNM (Gaussian network model), and compute the NMA (normal modes analysis) decomposition of the protein, using the implementation from the R Bio3D package \cite{grant2006bio3d}. Each of the $k$ modes represents a separate direction of motion, from large-scale motions (the smallest eigenvalues) to the high-frequency vibrations of hydrogen atoms (see Section~\\ref\{sec:nma\} for more details and the mathematical derivation of the NMA). Each mode corresponds to a different level in our hierarchical representation.

Hinge residues are obtained in a similar fashion to that demonstrated by HingeProt\cite{emekli2008hingeprot}, as follows. For each mode, $i$, we compute the mean square fluctuation matrix, $[F]_i$ (see Supplemental Information for the derivation):
\begin{equation}
 [F]_i = \frac{3k_BT}{\gamma}\lambda_i^{-1}\mathbf{u}_i\mathbf{u}_i^\intercal,\label{eqn:cc_mat_inline}
\end{equation}
where $\lambda_i$ and $\mathbf{u}_i$ is the eigenvalue and eigenvector of mode $i$, and $k_B$, $T$, and $\gamma$ are the Boltzmann constant, temperature, and uniform force constant, respectively. Regions of this matrix with the same sign form the rigid domains, and individual residues where the sign changes (from positive to negative) become the hinge residues. For practical purposes, we collapse domains with only a small number of residues.

\begin{figure}
  \centering
  \begin{minipage}[c]{0.27\linewidth}
    \centering
    \includegraphics[width=\linewidth]{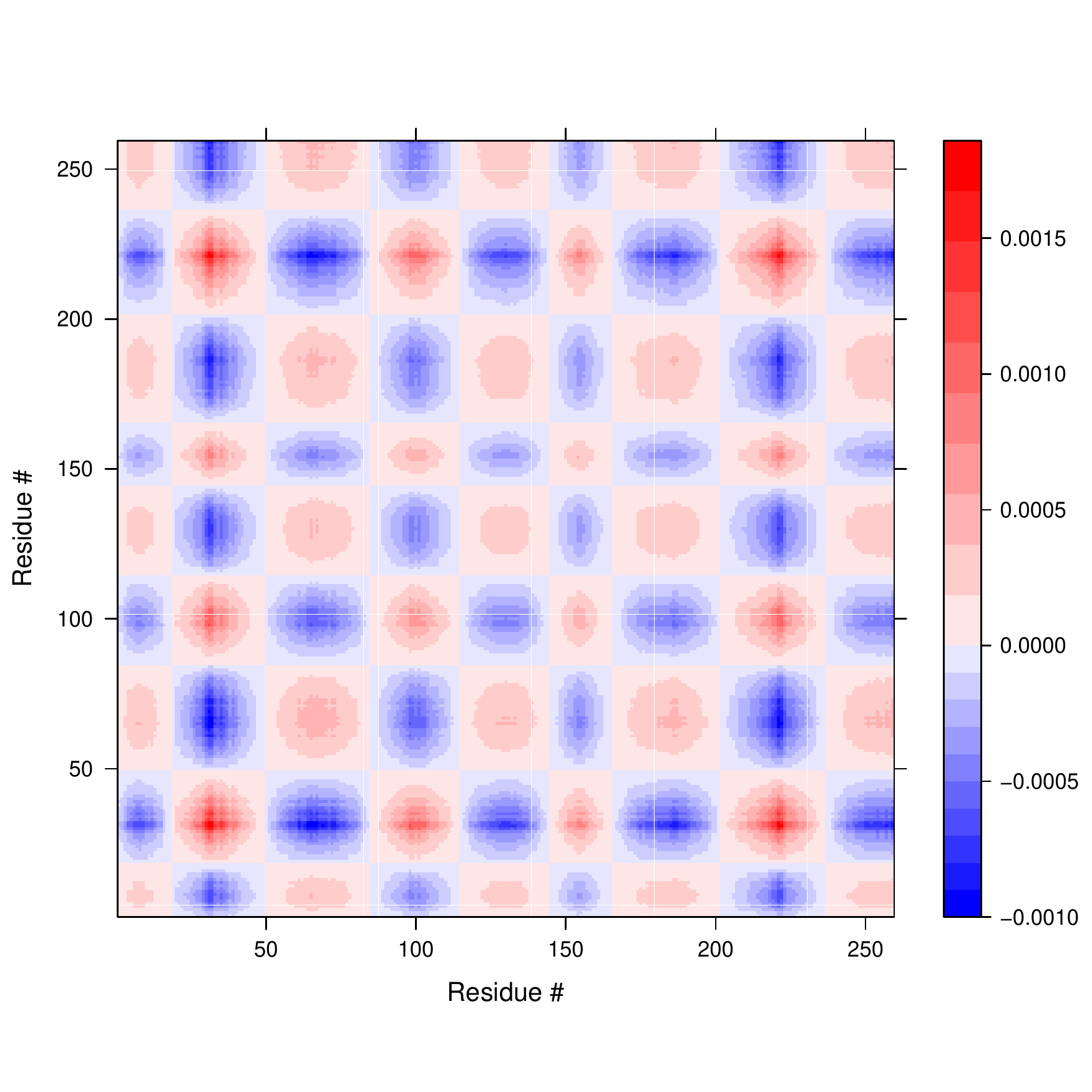}
  \end{minipage}%
  \begin{minipage}[c]{0.24\linewidth}
    \centering
    \includegraphics[width=\linewidth]{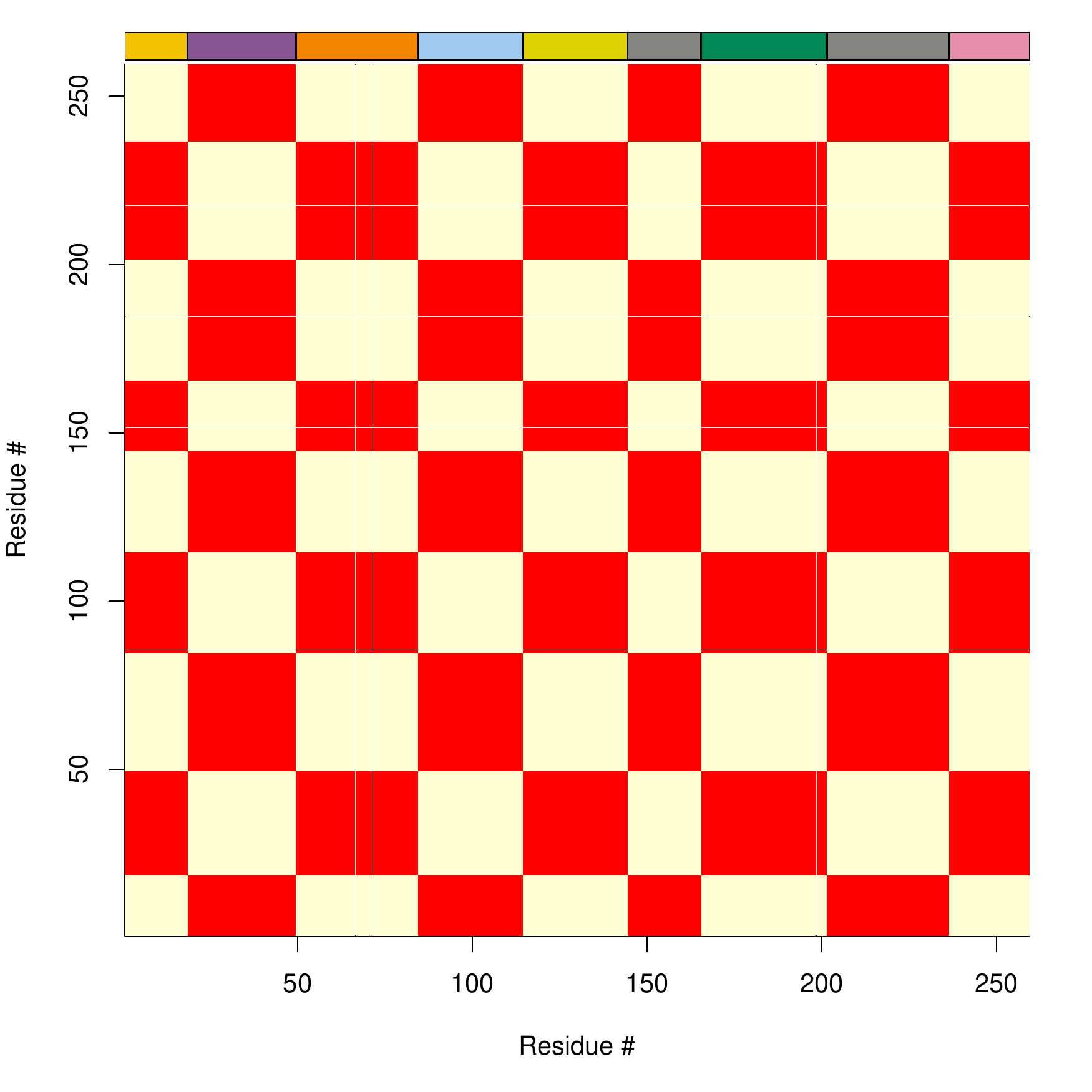}
  \end{minipage}%
  \begin{minipage}[c]{0.28\linewidth}
    \centering
    \includegraphics[width=\linewidth]{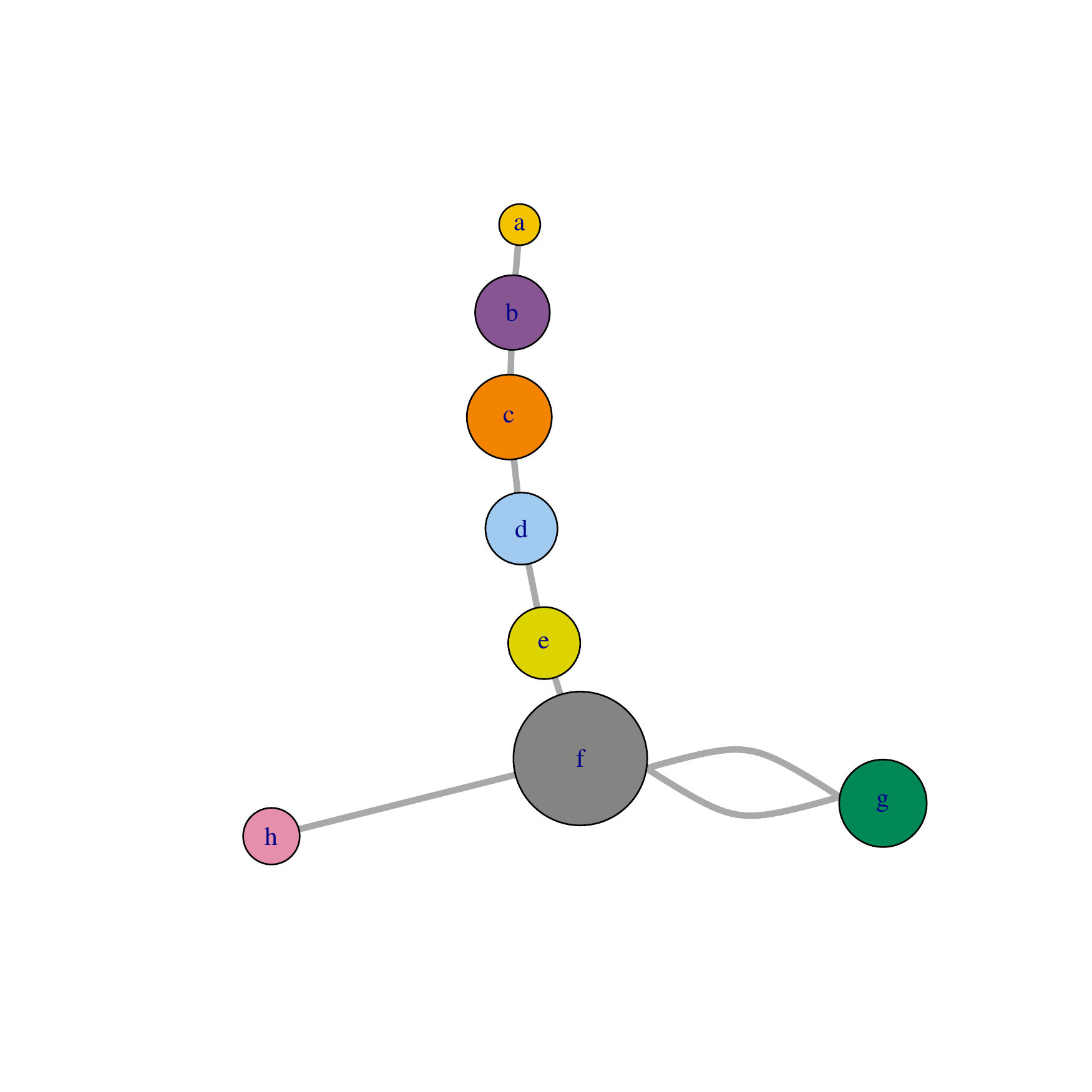}
  \end{minipage}%
  \begin{minipage}[c]{0.23\linewidth}
    \centering
    \hspace{-1cm}
    \includegraphics[width=\linewidth]{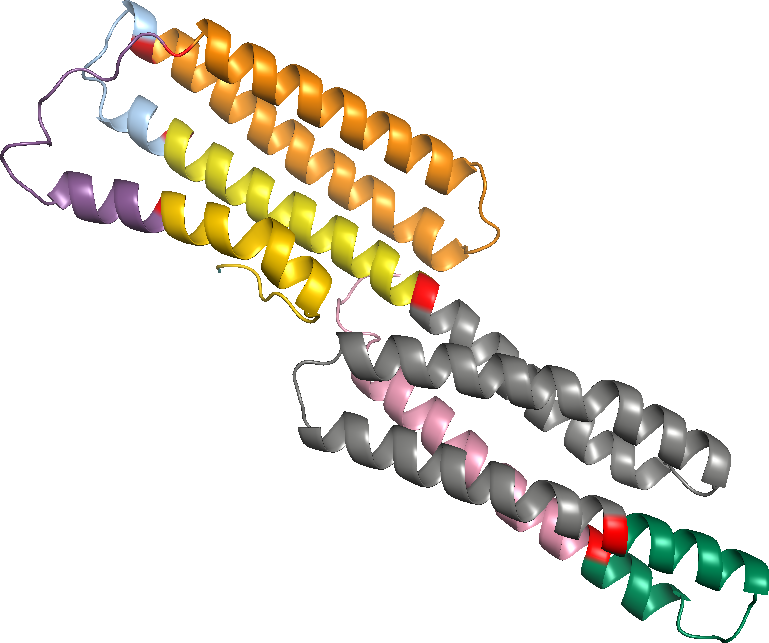}
  \end{minipage}%
  \caption{The NMA decomposition of 1RKE receptor, for the second non-trivial mode. From left to right: the cross-correlation fluctuation matrix, $[F]_2$; the sign of entries of $[F]_2$, with short domains removed; the domain graph representation of the protein, where the size of each node represents the number of residues in that domain; and the domain graph representation mapped onto the 3d structure of the protein, colored according to domain with hinge residues colored red. Hinges that are also flexible connectors separate all domains but $f$ (gray) and $g$ (green), which are connected by hinges that would not form a cut in the domain graph representation.}\label{fig:mode_decomp}
\end{figure}

The final stage at a given level is to construct a domain graph representation of the protein at that level, where nodes in the graph represent the rigid domains, and edges in the graph are inserted wherever two domains are in contact with each other, \ie any atom from one rigid domain is within $r_c$ of any atom from another domain (see Figure~\ref{fig:mode_decomp} for the decomposition a single level in the hierarchy). From this graph, we categorize each hinge as {\it flexible connector} if the removal of the hinge residue would form a cut of the domain graph representation, \ie its removal would result in two disjoint subgraphs.

Once we have obtained the domain graph representation of the protein for each of the $k$ NMA modes, we construct a multi-graph of the domain hierarchy for the entire protein \cite{F3Dock}. At the top level of the hierarchy are the hinges and domains computed by the first non-trivial mode (\ie with the smallest eigenvalue), representing more broad, global motions. The next level of the hierarchy is represented by the second smallest eigenvalue, and so on until all $k$ modes have been used. We also assign a weight, $w_k$, to all hinges at level $k$ of the hierarchy, arising from Equation~\ref{eqn:cc_mat_inline}:
\begin{equation}
  w_k = 3k_BT\lambda_k^{-1}.\label{eqn:hinge-weight}
\end{equation}

The final dimension of the product space of sampling is then $K_R+K_L$, where $K_R$ ($K_L$) is the number of hinges from all $k$ levels for the receptor (ligand), creating a product space of $SO(2)^{K_R+K_L}$ ($SO(2)$ is the special orthogonal group of rotations about a fixed axis). For the dataset used in this paper, the value of $K_R+K_L$ range from 21 hinge residues (3FN1) to 70 (1BKD). It is well known that generating a small number of good (\ie low discrepancy) samples is difficult in high dimensions, so to overcome this issue, we use the low-discrepancy sampling protocol developed by \cite{BBCZ_2014} when generating samples.


\subsubsection{Ramachandran distributions of amino acids}

Generating a low-discrepancy sampling of a high-dimensional space still requires a large number of samples to completely cover the product space. However, most of these samples will lead to physically impossible protein structures: clashes between nearby atoms, steric strain, or even a protein that is no longer biologically active. We would like to reduce the sampling space for a given torsion angle from all of $SO(2)$ to only the relevant, low-energy regions. 

\begin{figure}
 \centering
 \includegraphics[width=0.3\textwidth]{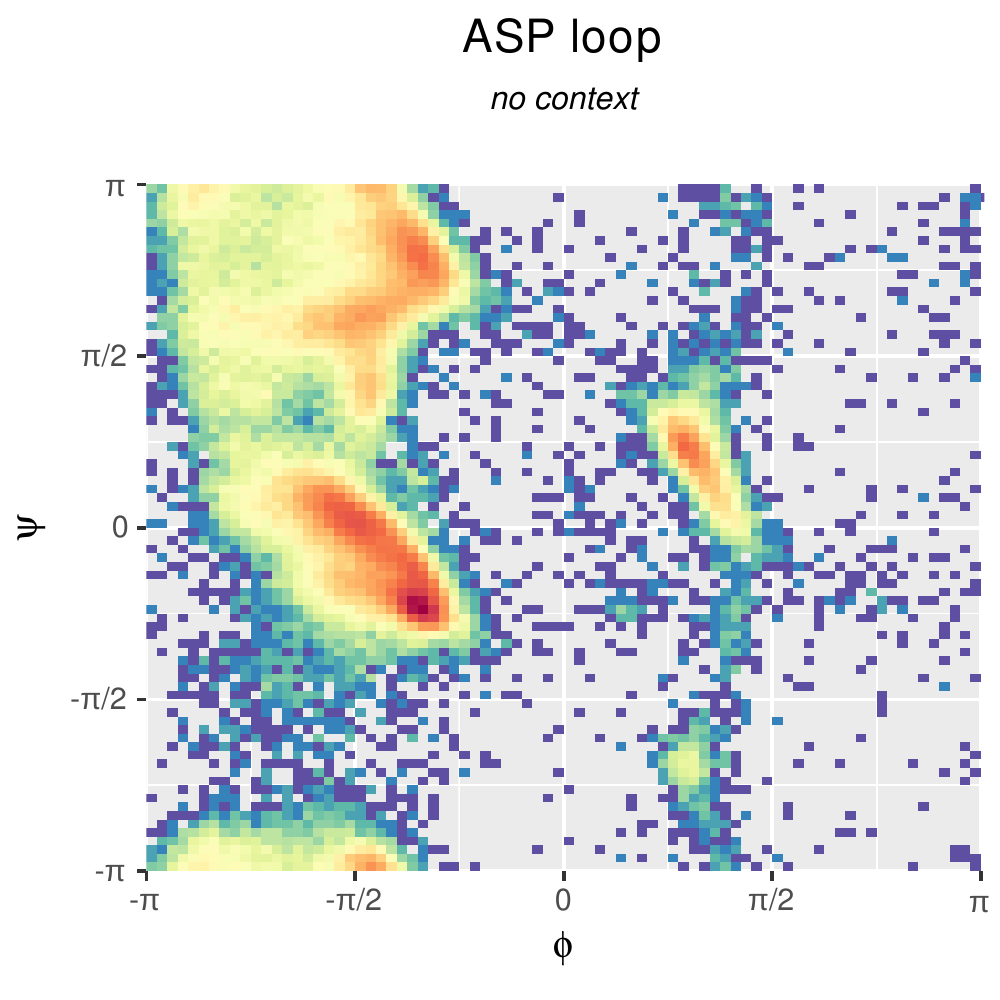}
 \includegraphics[width=0.3\textwidth]{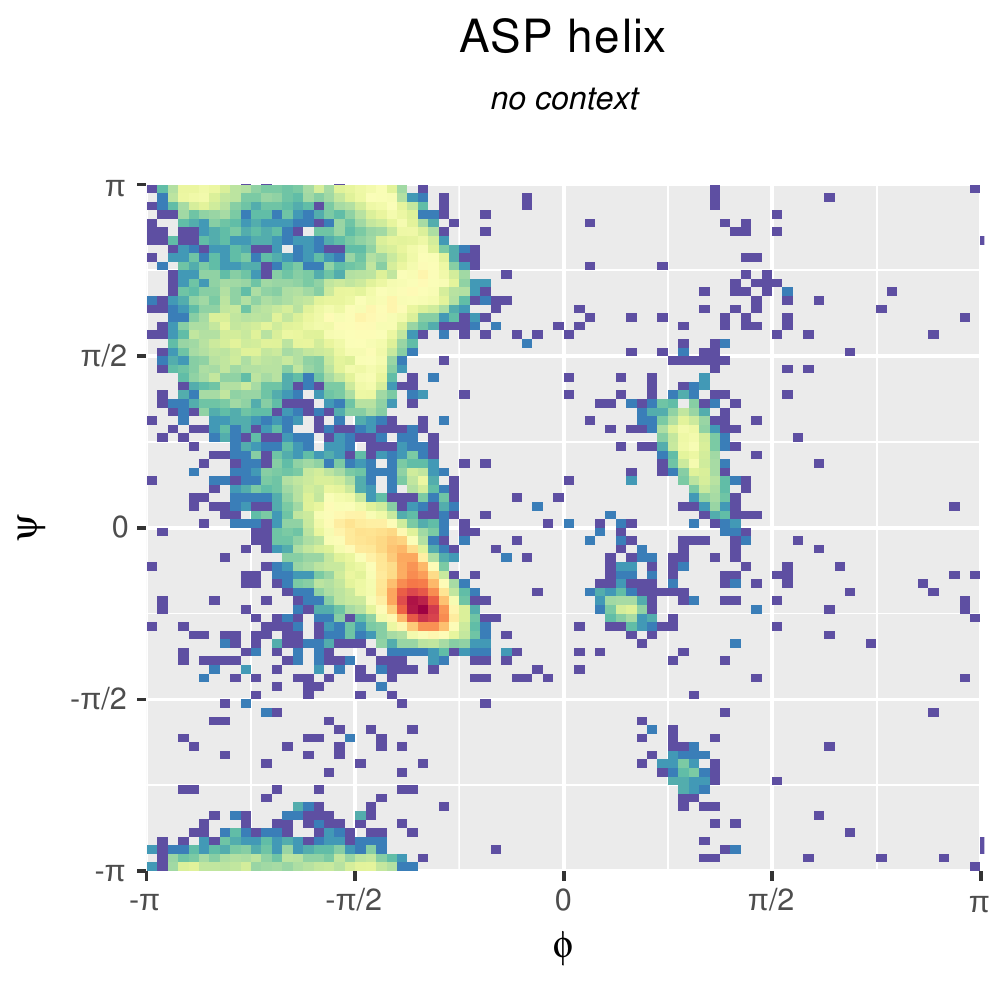}
 \includegraphics[width=0.3\textwidth]{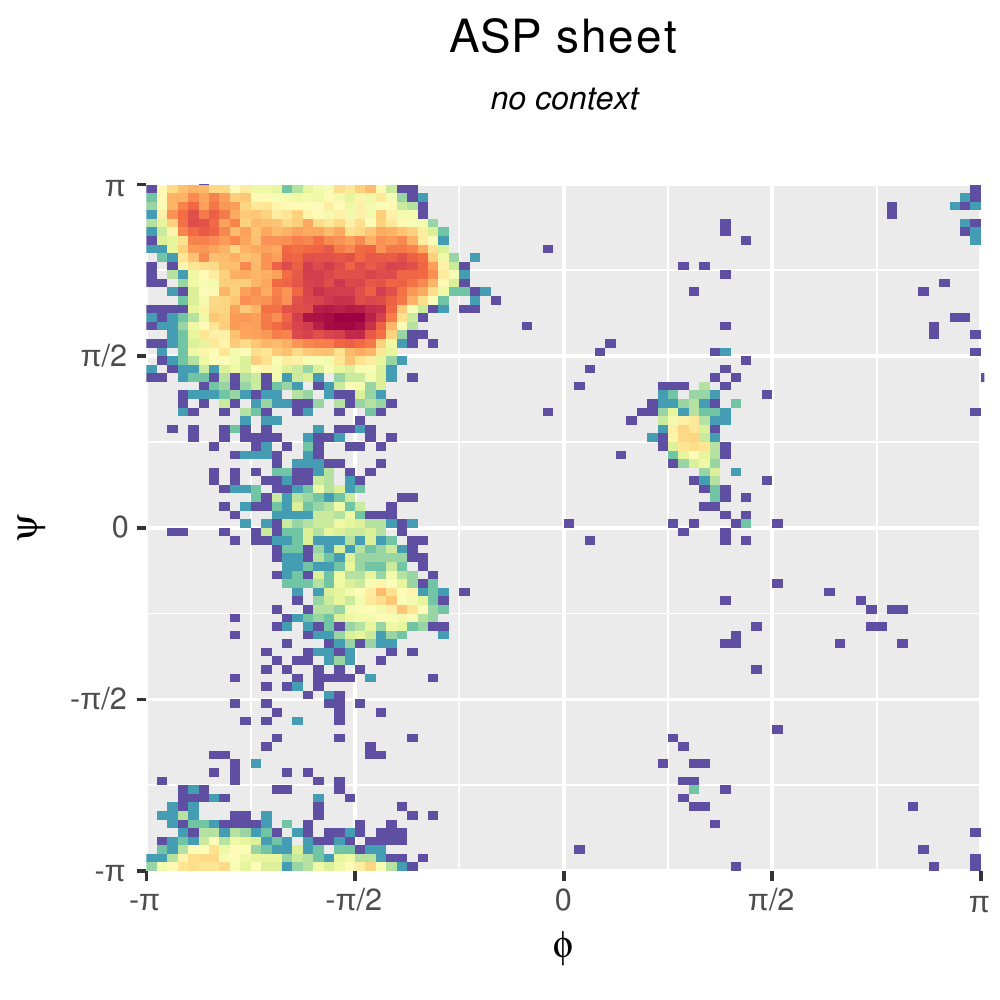} \\

 \includegraphics[width=0.3\textwidth]{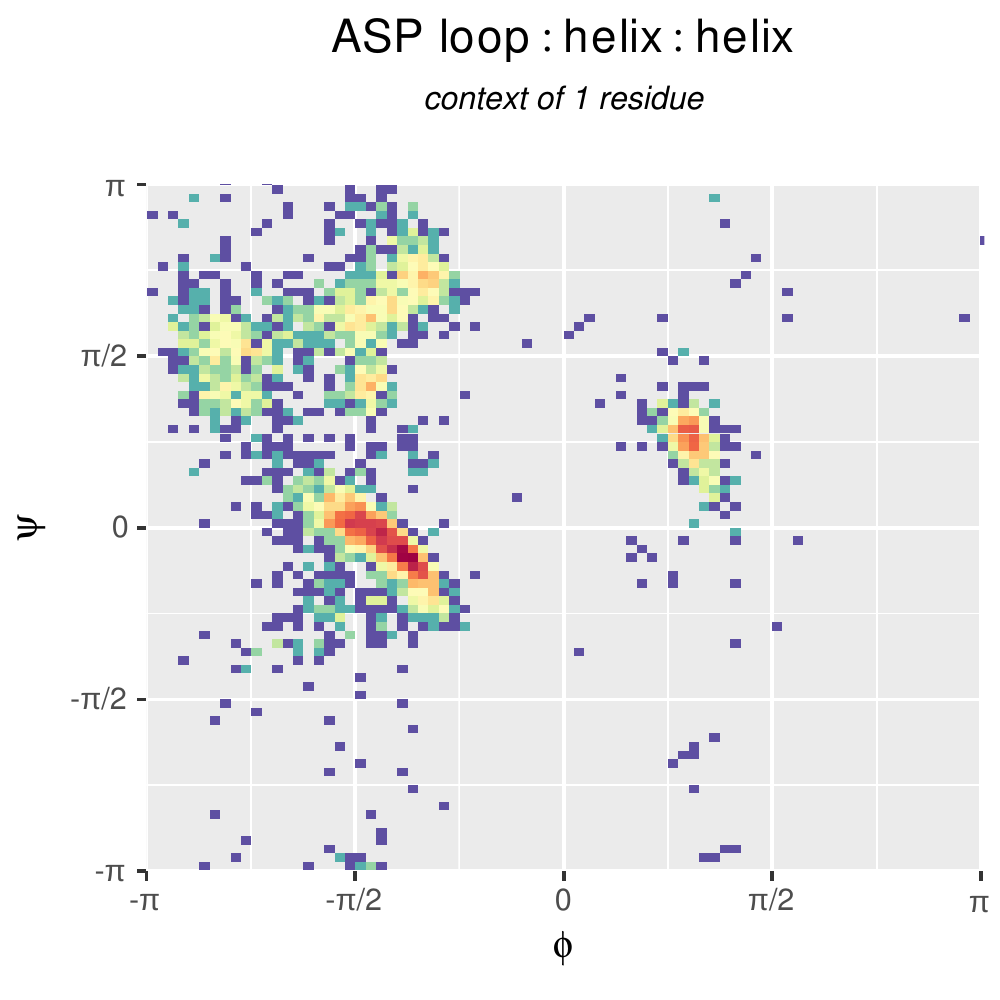}
 \includegraphics[width=0.3\textwidth]{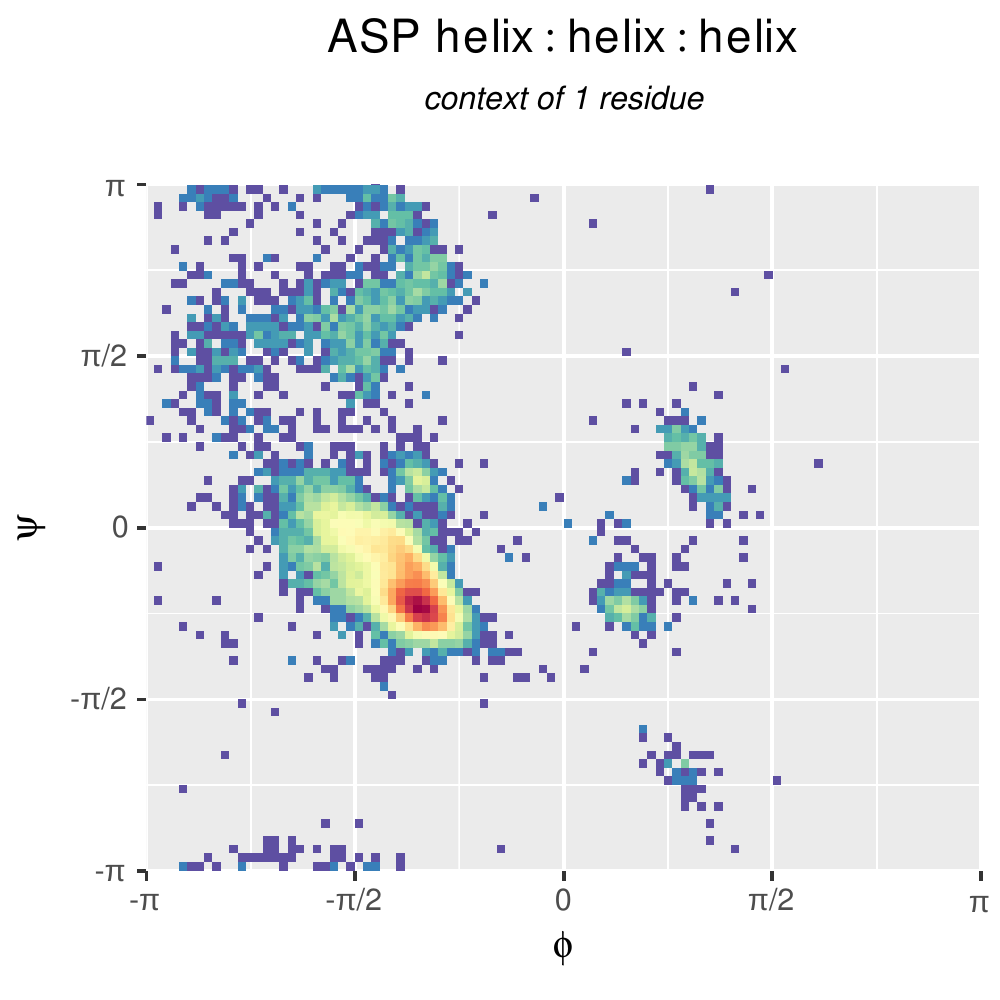}
 \includegraphics[width=0.3\textwidth]{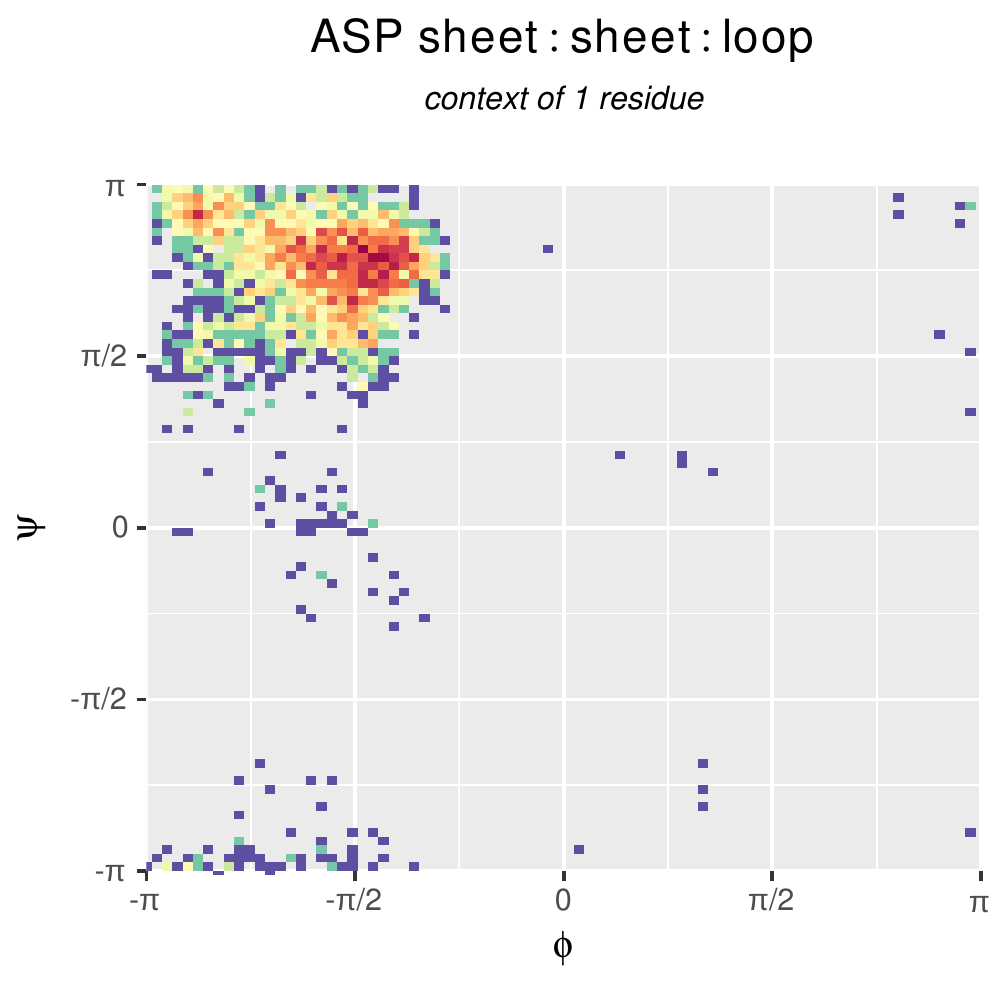} \\
 \caption{Ramachandran distributions for aspartic acid under different parameterizations. {\it Top row}: distributions where $ss_i$ is a loop, helix, and sheet (respectively). {\it Bottom row}: distributions where $ss_{i-1}$, $ss_i$, and $ss_{i+1}$ are respectively loop-helix-helix, helix-helix-helix, and sheet-sheet-sheet. Note that the distributions are more tightly clustered with the gain of additional context.}\label{fig:rama_distribs}
\end{figure}

In this work, we generate neighbor-dependent Ramachandran probability distributions from a set of $\sim$15k high-quality, non-homologous protein structures. These structures were obtained from the Pisces server \cite{wang2003pisces} with the following parameters: $<50\%$ sequence identity, X-ray crystallography resolution $<2.0$, and R-factor value of $<0.25$. From this set, we generate discrete probability distributions for each backbone torsion angle pair, conditioned on the amino acid type and secondary structure type of the previous and following residues. In other words, the probability of a given $(\phi,\psi)$ backbone torsion pair is given by:
\begin{equation}
  \Pr\left(\phi_i, \psi_i\right) = \Pr\left(\phi_i, \psi_i | ss_{i-1}, ss_i, ss_{i+1}, aa_i\right),\label{eqn:nd-rama}
\end{equation}
where $ss_i$ and $aa_i$ are the secondary structure and amino acid types of residue $i$, repsectively. Figure~\ref{fig:rama_distribs} shows the distributions for aspartic acid.

To generate samples of a given protein, we would like to draw samples for each hinge from the neighbor-dependent Ramachandran distributions. However, we also recognize that the input protein has important structural elements that should be preserved. For this reason, we convolve the discrete Ramachandran distribution with a bivariate von Mises distribution (the two-dimensional variant of the approximately-Gaussian distribution on a unit circle, \eg $[-\pi, \pi)^2$ \cite{mardia_vonmises}), centered at the given backbone torsion angle. The cosine variant of the bivariate von Mises distribution is given as follows:
\begin{equation}
\Pr\left(\phi,\psi\right) = Z_c(\kappa_1,\kappa_2,\kappa_3)\exp\left(\kappa_1\cos\left(\phi-\mu\right) + \kappa_2\cos\left(\psi-\nu\right)+\kappa_3\cos\left(\phi-\mu-\psi+\nu\right)\right),
\end{equation}
where $\mu$ and $\nu$ describe the mean for $\phi$ and $\psi$, $\kappa_1$ and $\kappa_2$ describe their concentration, and $\kappa_3$ describes their correlation. If $\kappa_3$ is zero and $\kappa_1=\kappa_2=s$, then $s$ can be used to increase or decrease the amount of bias the input structure has on the Ramachandran distributions. Lower values of $s$ (lower concentration) bias more toward the general Ramachandran distributions, while higher values of $s$ bias more towards the input protein structure.

To generate a specific $(\phi,\psi)$ pair, we use inverse CDF sampling: construct the discrete CDF of the von Mises-convolved Ramachandran distribution, then draw a single random sample from $[0,1)$, which serves as an index into the CDF to obtain the corresponding values of $\phi$ and $\psi$.

With the internal angles representation and hierarchical decomposition of the protein as input, we perform the following importance sampling protocol on each level, $l$:
\begin{enumerate}
  \item For each hinge at level $l$, $h_j^{(l)}$, let $i$ be the index of the residue corresponding to this hinge.
      \begin{enumerate}
          \item Generate the pair $(\hat\phi, \hat\psi)$, drawn from the von Mises-convolved neighbor-dependent Ramachandran distributions
          \item Let $\rho_l$ be the probability of a given hinge residue changing, arising from $w_l$ in Equation~\ref{eqn:hinge-weight}: $\rho_l=\min(1, w_l)$
          \item if $h_j^{(l)}$ is a cut or no other non-cut hinges have been sampled at level $l$, set $(\phi_i, \psi_i)$ to $(\hat \phi, \hat \psi)$ with probability $\rho_l$; otherwise, keep the original $(\phi_i, \psi_i)$ pair
      \end{enumerate}
    \item From the internal angle sample, generate the explicit structure in $\mathbb{R}^3$
    \item Compute the number of clashes caused by hinges at level $l$, and accept the torsion angle changes for level $l$ if the number of clashes are less than some parameter $c$. We define a {\it clash} as two atoms occupying the same space in $\mathbb{R}^3$.
\end{enumerate}

As this protocol only generates the backbone atoms, we then use SCWRL4\cite{Krivov2009improved} to add side-chain atoms, followed by a brief energy minimization step with Amber16\cite{amber2016}. Finally, we rank each sample by free energy, and keep only the samples with the lowest energy. These final two steps (minimization and ranking by energy) prevent from using samples that are biologically irrelevant.

\subsection{Benchmark dataset}
In this research, we are interested in 1) modeling the uncertainty of a given protein-protein docking algorithm, but also 2) improving the existing docking results in the unbound-unbound case. For this reason, we are interested in the performance of single protein pairs. The Zlab benchmark 5\cite{zlab5} contains a set of proteins that have had the X-ray structure determined both in isolation and together, and consist of 254 protein pairs classified as either {\it difficult}, {\it medium difficulty}, or {\it rigid-body}, depending on the interface RMSD (iRMSD). The difficult class of proteins have an iRMSD of $>2.2$\AA, which means there is typically some movement between bound and unbound conformations.

To select our set of input structures, we docked each protein classified as ``difficult'' in both the bound and unbound conformations with F2Dock \cite{f2dock,F2DockGBRerank}, a rigid-body docking algorithm (see Supplemental Information, Figure~\\ref\{fig:f2dock\_results\}). We selected those proteins that performed well in the bound structure but poorly with the unbound structure as candidates in our benchmark. The criteria we use for differentiating between success and failure is whether there exists a single docked conformation in the top 1000 reported poses with an iRMSD within 5\AA of the actual bound conformation.

Since we are primarily interested in the single-body docking problem (and not the multi-body docking problem), we only kept the single-chain proteins for our experiment, which led to 10 single-chain proteins that perform well when using the bound conformation but not when unbound. In addition, we also included the 3 single-chain proteins that performed poorly when the both the bound and unbound conformation were used. Statistics on the size and free energy of each protein are given in Table~\ref{tab:benchmark_stats}.
\begin{table}
  \centering
  \caption{Protein structures used in dataset, labeled according to the ID from the ZLab benchmark 5 \cite{zlab5}. The top section contain those that performed well when bound, the bottom section containing those that did not.}\label{tab:benchmark_stats}
  \begin{tabular}{l|rrr|r}
          & \multicolumn{3}{c|}{\# Residues} & \\
      ID  & receptor & ligand & contact & $\Delta$Energy (J) \\\hline
     1ATN & 372      & 258    &  36 & 131 \\
     1F6M & 320      & 108    &  62 &  87 \\
     1FQ1 & 183      & 295    &  53 & 367 \\
     1BKD & 439      & 166    &  97 & 425 \\
     1R8S & 160      & 187    &  61 & 439 \\
     1RKE & 262      & 176    &  68 & 524 \\
     1ZLI & 306      &  74    &  77 & 212 \\
     2C0L & 292      & 122    &  92 & 366 \\
     2I9B & 265      & 122    & 101 & 387 \\
     2J7P & 292      & 265    &  80 & 370 \\
     2OT3 & 253      & 157    &  69 & 428 \\
     3FN1 & 160      &  90    &  38 & 315 \\\hline\hline

     1H1V & 368      & 327    &  74 & 180 \\
     1Y64 & 411      & 357    &  66 & 192 \\
     3AAD & 264      & 153    &  42 &  66 \\
  \end{tabular}
\end{table}

\subsection{F2Dock results on ZLab benchmark 5}
\begin{figure}[ht]
 \centering
  \includegraphics[width=\linewidth]{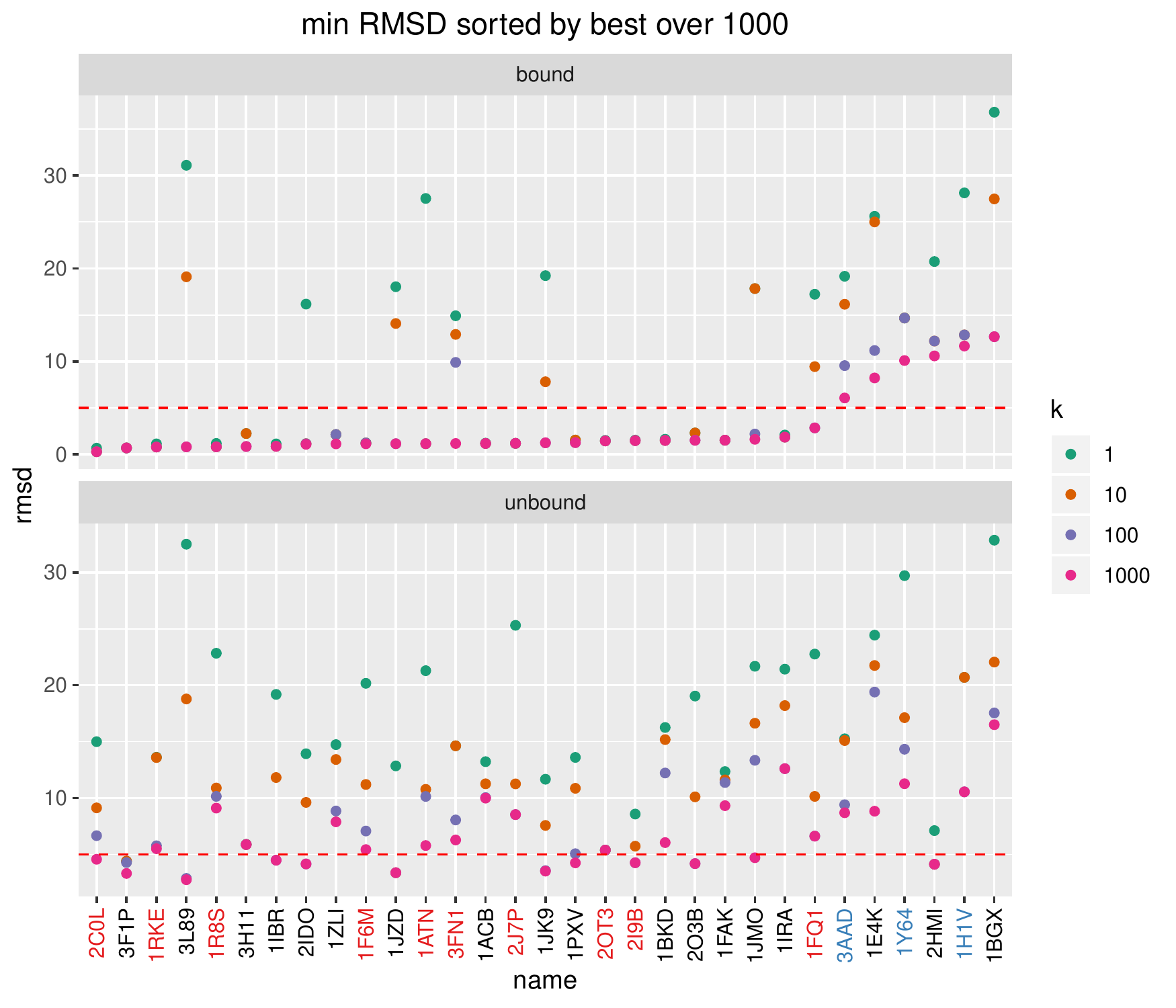}
  \caption{Results from F2Dock on ZLab benchmark 5 \cite{zlab5}, labeled ``difficult,'' when bound (top) and unbound (bottom) structures were used. The structures with names in blue are those that performed well in the bound conformation and poorly when unbound, and those with names in red performed poorly both with unbound and bound. Both the blue and the red structures were used as the benchmark in this paper.}\label{fig:f2dock_results}
\end{figure}

Figure~\ref{fig:f2dock_results} shows the initial results from the ZLab benchmark 5 \cite{zlab5} on all the proteins labeled ``Difficult.'' For each bound and unbound protein pair, the interface RMSD to the true bound pair is computed for the top $k$ pairs. The minimum iRMSD for different values of $k$ are plotted, and proteins along the x-axis are sorted according to the minimum iRMSD among the top 1000 conformations. This highlights the differences in this rigid-body docking algorithm when the bound and unbound proteins are given: F2Dock is able to correctly identify a conformation within 5\AA for most of the bound inputs (and over half of the proteins report the top conformation within 5\AA), but most of the unbound inputs do not have any hits within the top 1000 conformations.

\section{Results and discussion}
\subsection{Conformational sampling distributions}
\begin{figure}
  \includegraphics[width=\linewidth]{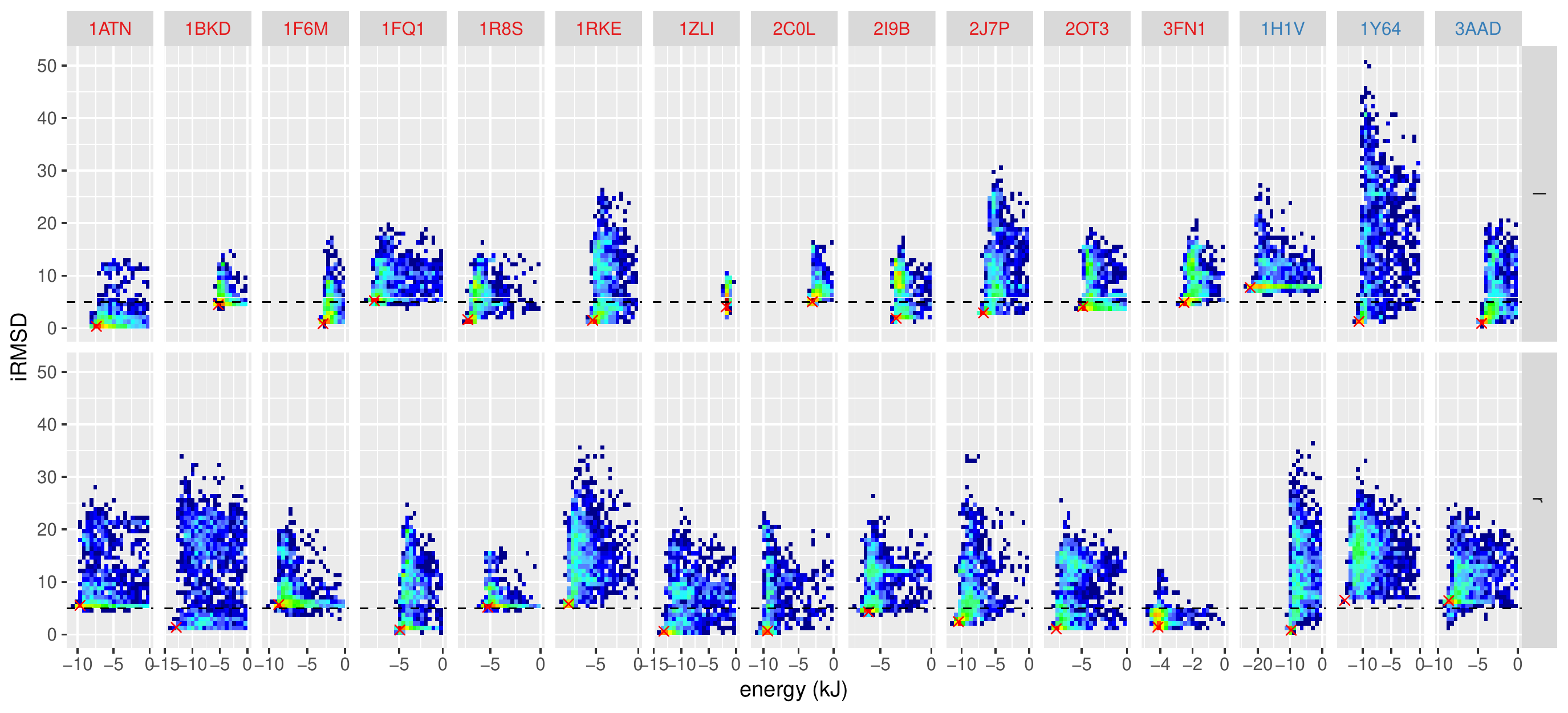}
  \caption{Plot of iRMSD (against the bound conformation) vs energy for individual protein pairs (receptor and ligand) for all proteins in this dataset. The black dotted line shows iRMSD$=5$, or the value at which a match is considered a ``hit,'' and the red ``X'' marks the spot of the original unbound protein. For all proteins, there exist some samples that improve on both the iRMSD and energy, and some of the proteins, such as 3AAD receptor and 1ZLI ligand, improve upon the iRMSD greatly. Strong convergence is shown by a funnel-shaped energy landscape, and is seen for many protein pairs. Protein labels are colored red (good when bound) and blue (bad when bound).}\label{fig:conf_distr}
\end{figure}

Our primary concern for generating a good set of low-discrepancy samples is to ensure that the samples cover a good portion of the feasible set of the protein conformational space. We consider two metrics for ranking samples: 1) the free energy of individual proteins, and 2) the iRMSD from the sample to the bound conformation. The first of these metrics is an unbiased measure of protein stability: if all samples have abnormally high energy, they are unlikely to be biologically feasible. For the second metric, it is possible that the bound conformation lies in an energy well that is made more available when in combination with the second protein pair. For this reason, we are also interested in knowing how close we can get to the bound conformation. Figure~\ref{fig:conf_distr} shows the energy vs iRMSD for 1000 samples of each protein.

\subsection{Improvements in unbound-unbound docking}
By generating a set of proteins that have a closer iRMSD to the bound conforation, we are able to improve on the blind unbound-unbound docking protocol, for both rigid-body and flexible docking algorithms. We compare the results for the bound and unbound case for F2Dock, Rosetta, and SwarmDock. We perform the bound-bound and unbound-unbound docking for each docking algorithm, and compute the iRMSD on the reported poses. For F2Dock and Rosetta, the number of reported poses is variable, which we set at 1000; SwarmDock reports a fixed number of results, so this number varies from 465--548 poses. Since F2Dock and Rosetta both have command-line interfaces, we also generate 50 samples of the unbound conformation of each protein and generate the top 100 poses for each of these. The minimum iRMSD for each protein (bound, unbound, and samples for F2Dock and Rosetta) are found in Table~\ref{tab:f3dock_res}.

\begin{table}
  \centering
  \caption{Best RMSD (over top 1000 poses for F2Dock and Rosetta and all poses for SwarmDock) for proteins included in this dataset. A single asterisk marks proteins that had at least one hit (iRMSD $<5$\AA) in the top poses. F2Dock and Rosetta statistics for sampled proteins are also included.}\label{tab:f3dock_res}
  \begin{tabular}{l|rrr|rrr|rr}
    & \multicolumn{3}{c|}{F2Dock} & \multicolumn{3}{c|}{Rosetta} & \multicolumn{2}{c}{SwarmDock}\\
    ID   & bound & unbound & sampled &  bound & unbound &sampled & bound & unbound \\\hline
    1ATN &  1.2  &     7.2 &   *4.6  &   0.08 &     8.6 &    6.8 &  0.98 &   *4.6  \\
    1BKD &  1.3  &     8.7 &   *4.9  &   0.23 &    16.9 &    5.4 &  0.68 &    8.7  \\
    1F6M &  1.2  &     8.1 &   *5.0  &   0.11 &    17.9 &   13.4 &  0.69 &    5.6  \\
    1FQ1 &  2.3  &     6.4 &   *4.4  &   0.48 &    15.0 &    8.2 &  3.55 &    5.6  \\
    1R8S &  1.7  &     9.4 &    6.2  &   0.22 &    14.5 &    6.2 &  0.72 &    5.1  \\
    1RKE &  0.8  &     7.1 &    5.8  &   0.15 &    15.1 &   12.8 &  0.65 &    5.4  \\
    1ZLI &  0.6  &    10.0 &   *4.6  &   0.13 &    10.6 &    7.1 &  0.71 &    9.0  \\
    2C0L &  0.5  &    *4.8 &   *4.0  &   0.30 &    12.2 &    8.2 &  0.75 &   *3.8  \\
    2I9B &  1.1  &     8.4 &   *4.7  &   0.09 &    13.6 &    8.8 &  7.93 &    6.5  \\
    2J7P &  1.4  &     7.2 &   *3.4  &   1.12 &    17.2 &   14.9 &  0.60 &    6.6  \\
    2OT3 &  1.1  &     5.1 &   *3.9  &   0.16 &    15.6 &    6.8 &  0.92 &    6.0  \\
    3FN1 &  1.0  &     5.5 &   *4.9  &   0.10 &     9.9 &   *4.8 &  0.53 &   *4.1  \\\hline\hline

    1H1V &  8.8  &    11.0 &    8.0  &   0.27 &    18.8 &   13.5 &  0.68 &    9.1  \\
    1Y64 &  9.3  &    11.4 &   10.7  &   1.9  &    35.0 &   15.6 &  1.37 &   11.7  \\
    3AAD &  7.0  &     8.2 &    5.3  &   0.39 &    22.0 &    9.2 &  2.36 &    7.1  \\
  \end{tabular}
\end{table}

The results from the iRMSD statistics suggest a few findings. First, the flexible algorithms (Rosetta and SwarmDock) are better at docking the bound-bound conformations than the rigid-body one (F2Dock). This is potentially due to the fact that clashes prevent the rigid body docking algorithm from correctly identifying the best conformation, but also could be due to the fact that each docking algorithm uses a different energy function, which is better tuned in the flexible algorithms. More interesting, however, is the observation that the rigid-body docking algorithm performs just as good as the flexible docking algorithms in the unbound-unbound case, possibly because of the high-dimensional landscape that must be traversed by the docking algorithm. In addition, this suggests that the input to the algorithm (e.g. unbound or bound) is an important characteristic of the docking run, and variations in input structure must be accounted for. The final important observation is that for each protein, the best iRMSD for each sample is always less than the iRMSD for the unbound-unbound case. This not only suggests that the sampling protocol is sound (leading to better results), but that there is further variance from input structures that must be described with empirical certificates.

\subsection{Probabilistic certificates from Quasi-Monti Carlo samples}
To describe the uncertainty of the results of the docking algorithm, we compute the probabilistic certificates arising from the Chernoff-like bounds of the sampled algorithms. This will provide a metric that can compare across proteins (for the same docking algorithm) and across docking algorithms (for the same protein, or over all proteins). We could provide probabilistic certificates for any QOI; however, we are primarily interested in bounding the binding free energy. If the reported free energy is tightly bound by a probabilistic certificate, we are more confident that we have identified the correct free energy.

\begin{figure}
  \centering
  \includegraphics[width=4in]{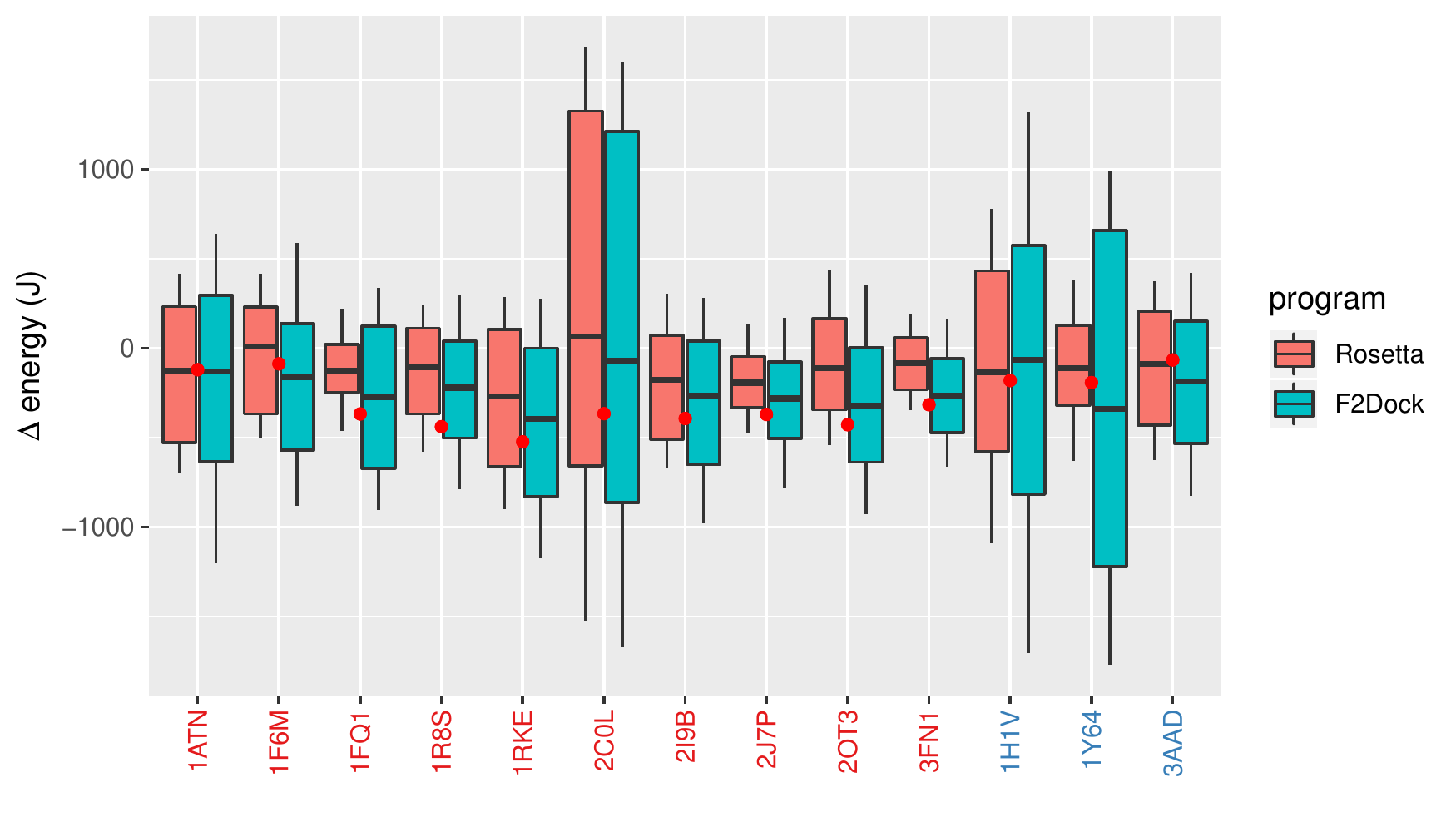}
  \caption{Probabilistic bounds compared to ground truth for values of $\Delta$G. Protein labels are colored red (good when bound) and blue (bad when bound). The box shows the value of the certificate at $\Pr=0.9$ and the tails show the min/max values, and the red point shows the true statistic, computed on the bound-bound form of the protein. Compare this plot with Supporting information Table~\\ref\{tab:chernoff\_den\}, where specific values of $t$ for the certificate are given.}\label{fig:certificate_den}
\end{figure}

Figure~\ref{fig:certificate_den} shows a comparison of the certificate for $\Delta$G of each protein (at $\Pr=0.9$), and includes the true QOI, computed on the bound-bound conformation. Explicit values of each QOI are included in Supplemental Tables~\\ref\{tab:chernoff\_den\} and \\ref\{tab:chernoff\_crmsd\}. For some of the proteins (\eg 3FN1), the provided certificate is much tighter than others (\eg 2C0L). This also allows us to directly compare the two different programs in terms of docking uncertainty. While the rigid F2Dock algorithm occasionally has higher bounds (see Table~\\ref\{tab:chernoff\_den\}, with high probability the true statistic lies within the $\Pr=0.9$ certificate range. The Rosetta results usually contain the true QOI; however, for some proteins (such as 3FN1), the true QOI does not even lie within the min/max range.

\subsection{Chernoff-like bounds for specific QOI}
Table~\ref{tab:chernoff_den} shows the Chernoff-like bounds for delta energy, where $t$ is an absolute value, \ie $\Pr(f(X)-E[f]|)>t$, and Table~\ref{tab:chernoff_crmsd} shows the Chernoff-like bounds for iRMSD, where $t$ is a fraction of the mean, \ie $\Pr(|f(X)-E[f]|>t*E[f])$. Bold-face values are those where the certificate probability is $>0.9$, or with high confidence.
\begin{table}[ht]
  \caption{Chernoff-like bounds for $\Delta E$ of each protein, for Rosetta and F3Dock. Values of $t$ (see Equation~\ref{eqn:chernoff}) are in terms of absolute value of $\Delta E$, \eg $\Pr(|f(X)-E[f]|)>t$. Boldface values are those values where the probability is greater than 0.9.}\label{tab:chernoff_den}
\centering
\begin{tabular}{lr|rr|rrrrrrr}
pdb & program & mean & sd & 0.005 & 0.01 & 0.05 & 0.1 & 0.2 & 0.3 & 0.5 \\
  \hline
1ATN & F2Dock  & -0.129 & 0.292 & {\bf 0.983} & {\bf 0.968} & 0.847 & 0.701 & 0.457 & 0.269 & 0.082 \\
     & Rosetta & -0.128 & 0.210 & {\bf 0.973} & {\bf 0.943} & 0.746 & 0.541 & 0.320 & 0.194 & 0.011 \\
\hline
1BKD & F2Dock  & -0.331 & 0.668 & {\bf 0.993} & {\bf 0.985} & {\bf 0.943} & 0.888 & 0.786 & 0.708 & 0.542 \\
     & Rosetta & -0.136 & 0.609 & {\bf 0.987} & {\bf 0.975} & 0.892 & 0.799 & 0.658 & 0.567 & 0.441 \\
\hline
1F6M & F2Dock  & -0.160 & 0.209 & {\bf 0.978} & {\bf 0.957} & 0.770 & 0.574 & 0.303 & 0.152 & 0.027 \\
     & Rosetta & 0.009 & 0.166 & {\bf 0.967} & {\bf 0.936} & 0.699 & 0.468 & 0.200 & 0.090 & 0.003 \\
\hline
1FQ1 & F2Dock  & -0.274 & 0.204 & {\bf 0.974} & {\bf 0.946} & 0.726 & 0.494 & 0.226 & 0.138 & 0.042 \\
     & Rosetta & -0.126 & 0.087 & {\bf 0.941} & 0.886 & 0.474 & 0.183 & 0.046 & 0.007 & 0.000 \\
\hline
1R8S & F2Dock  & -0.221 & 0.165 & {\bf 0.977} & {\bf 0.954} & 0.757 & 0.533 & 0.220 & 0.067 & 0.004 \\
     & Rosetta & -0.104 & 0.135 & {\bf 0.964} & {\bf 0.928} & 0.650 & 0.415 & 0.138 & 0.041 & 0.000 \\
\hline
1RKE & F2Dock  & -0.397 & 0.245 & {\bf 0.981} & {\bf 0.963} & 0.815 & 0.661 & 0.421 & 0.245 & 0.034 \\
     & Rosetta & -0.270 & 0.216 & {\bf 0.978} & {\bf 0.959} & 0.776 & 0.581 & 0.343 & 0.208 & 0.007 \\
\hline
1ZLI & F2Dock  & -0.229 & 0.194 & {\bf 0.979} & {\bf 0.954} & 0.754 & 0.539 & 0.279 & 0.149 & 0.010 \\
     & Rosetta & -0.083 & 0.150 & {\bf 0.954} & {\bf 0.911} & 0.583 & 0.359 & 0.200 & 0.091 & 0.000 \\
\hline
2C0L & F2Dock  & -0.070 & 0.623 & {\bf 0.996} & {\bf 0.989} & {\bf 0.942} & 0.888 & 0.752 & 0.610 & 0.382 \\
     & Rosetta & 0.066 & 0.580 & {\bf 0.997} & {\bf 0.993} & {\bf 0.962} & {\bf 0.927} & 0.786 & 0.560 & 0.294 \\
\hline
2I9B & F2Dock  & -0.268 & 0.214 & {\bf 0.984} & {\bf 0.969} & 0.844 & 0.674 & 0.377 & 0.163 & 0.010 \\
     & Rosetta & -0.177 & 0.176 & {\bf 0.971} & {\bf 0.945} & 0.733 & 0.509 & 0.274 & 0.102 & 0.000 \\
\hline
2J7P & F2Dock  & -0.282 & 0.129 & {\bf 0.965} & {\bf 0.932} & 0.669 & 0.417 & 0.121 & 0.024 & 0.000 \\
     & Rosetta & -0.193 & 0.089 & {\bf 0.952} & {\bf 0.902} & 0.555 & 0.244 & 0.030 & 0.004 & 0.000 \\
\hline
2OT3 & F2Dock  & -0.322 & 0.194 & {\bf 0.980} & {\bf 0.955} & 0.784 & 0.591 & 0.295 & 0.122 & 0.013 \\
     & Rosetta & -0.112 & 0.156 & {\bf 0.962} & {\bf 0.928} & 0.666 & 0.453 & 0.200 & 0.057 & 0.006 \\
\hline
3FN1 & F2Dock  & -0.267 & 0.128 & {\bf 0.966} & {\bf 0.936} & 0.692 & 0.416 & 0.111 & 0.026 & 0.000 \\
     & Rosetta & -0.083 & 0.088 & {\bf 0.962} & {\bf 0.906} & 0.560 & 0.266 & 0.021 & 0.000 & 0.000 \\
\hline\hline
1H1V & F2Dock  & -0.064 & 0.436 & {\bf 0.989} & {\bf 0.977} & 0.898 & 0.797 & 0.626 & 0.461 & 0.227 \\
     & Rosetta & -0.134 & 0.301 & {\bf 0.985} & {\bf 0.969} & 0.837 & 0.680 & 0.416 & 0.262 & 0.128 \\
\hline
1Y64 & F2Dock  & -0.339 & 0.541 & {\bf 0.993} & {\bf 0.984} & {\bf 0.917} & 0.841 & 0.687 & 0.536 & 0.353 \\
     & Rosetta & -0.110 & 0.140 & {\bf 0.961} & {\bf 0.925} & 0.656 & 0.376 & 0.125 & 0.058 & 0.004 \\
\hline
3AAD & F2Dock  & -0.187 & 0.199 & {\bf 0.977} & {\bf 0.957} & 0.783 & 0.594 & 0.303 & 0.144 & 0.010 \\
     & Rosetta & -0.088 & 0.172 & {\bf 0.968} & {\bf 0.938} & 0.700 & 0.459 & 0.228 & 0.117 & 0.002 \\
\end{tabular}
\end{table}

\begin{table}[ht]
  \caption{Chernoff-like bounds for iRMSD of each protein, for Rosetta and F3Dock. Values of $t$ (see Equation~\ref{eqn:chernoff}) are in terms of fraction of mean iRMSD, \eg $\Pr(|f(X)-E[f]|>t*E[f])$. Boldface values are those values where the probability is greater than 0.9.}\label{tab:chernoff_crmsd}
  \centering
  \begin{tabular}{lr|rr|rrrrrrr}
  \hline
  pdb & program & mean & sd & 0.005 & 0.01 & 0.05 & 0.1 & 0.2 & 0.3 & 0.5 \\
  \hline
1ATN & F2Dock  & 20.721 & 5.854 & {\bf 0.997} & {\bf 0.985} & {\bf 0.969} & {\bf 0.940} & {\bf 0.910} & 0.852 & 0.712 \\
     & Rosetta & 18.522 & 2.886 & {\bf 0.996} & {\bf 0.980} & {\bf 0.961} & {\bf 0.923} & 0.877 & 0.759 & 0.466 \\
\hline
1BKD & F2Dock  & 26.527 & 5.237 & {\bf 0.995} & {\bf 0.979} & {\bf 0.957} & {\bf 0.913} & 0.869 & 0.784 & 0.580 \\
     & Rosetta & 23.986 & 5.920 & {\bf 0.997} & {\bf 0.983} & {\bf 0.967} & {\bf 0.934} & {\bf 0.903} & 0.840 & 0.685 \\
\hline
1F6M & F2Dock  & 16.716 & 3.119 & {\bf 0.995} & {\bf 0.977} & {\bf 0.955} & {\bf 0.909} & 0.861 & 0.769 & 0.547 \\
     & Rosetta & 25.703 & 2.015 & {\bf 0.990} & {\bf 0.950} & 0.898 & 0.796 & 0.694 & 0.505 & 0.177 \\
\hline
1FQ1 & F2Dock  & 18.953 & 4.294 & {\bf 0.996} & {\bf 0.983} & {\bf 0.966} & {\bf 0.932} & 0.899 & 0.834 & 0.673 \\
     & Rosetta & 20.984 & 2.196 & {\bf 0.992} & {\bf 0.960} & {\bf 0.918} & 0.837 & 0.757 & 0.605 & 0.291 \\
\hline
1R8S & F2Dock  & 22.340 & 4.558 & {\bf 0.996} & {\bf 0.983} & {\bf 0.965} & {\bf 0.929} & 0.895 & 0.821 & 0.637 \\
     & Rosetta & 18.420 & 1.471 & {\bf 0.989} & {\bf 0.946} & 0.891 & 0.778 & 0.671 & 0.474 & 0.144 \\
\hline
1RKE & F2Dock  & 21.246 & 4.767 & {\bf 0.996} & {\bf 0.981} & {\bf 0.963} & {\bf 0.926} & 0.891 & 0.818 & 0.647 \\
     & Rosetta & 23.270 & 2.994 & {\bf 0.994} & {\bf 0.970} & {\bf 0.942} & 0.880 & 0.818 & 0.686 & 0.418 \\
\hline
1ZLI & F2Dock  & 16.633 & 3.034 & {\bf 0.996} & {\bf 0.979} & {\bf 0.959} & {\bf 0.920} & 0.879 & 0.797 & 0.601 \\
     & Rosetta & 20.167 & 2.305 & {\bf 0.993} & {\bf 0.963} & {\bf 0.928} & 0.855 & 0.784 & 0.647 & 0.349 \\
\hline
2C0L & F2Dock  & 16.846 & 4.626 & {\bf 0.997} & {\bf 0.988} & {\bf 0.976} & {\bf 0.953} & {\bf 0.929} & 0.882 & 0.766 \\
     & Rosetta & 18.800 & 4.188 & {\bf 0.996} & {\bf 0.980} & {\bf 0.961} & {\bf 0.921} & 0.882 & 0.804 & 0.631 \\
\hline
2I9B & F2Dock  & 18.675 & 4.699 & {\bf 0.997} & {\bf 0.986} & {\bf 0.971} & {\bf 0.941} & {\bf 0.909} & 0.851 & 0.708 \\
     & Rosetta & 21.238 & 3.441 & {\bf 0.994} & {\bf 0.972} & {\bf 0.943} & 0.886 & 0.830 & 0.726 & 0.492 \\
\hline
2J7P & F2Dock  & 22.141 & 4.723 & {\bf 0.996} & {\bf 0.980} & {\bf 0.960} & {\bf 0.920} & 0.882 & 0.806 & 0.626 \\
     & Rosetta & 30.311 & 3.905 & {\bf 0.996} & {\bf 0.977} & {\bf 0.954} & {\bf 0.905} & 0.856 & 0.750 & 0.482 \\
\hline
2OT3 & F2Dock  & 18.916 & 4.816 & {\bf 0.997} & {\bf 0.985} & {\bf 0.972} & {\bf 0.943} & {\bf 0.916} & 0.860 & 0.734 \\
     & Rosetta & 20.114 & 2.616 & {\bf 0.995} & {\bf 0.975} & {\bf 0.948} & 0.893 & 0.840 & 0.728 & 0.453 \\
\hline
3FN1 & F2Dock  & 18.234 & 3.301 & {\bf 0.996} & {\bf 0.979} & {\bf 0.957} & {\bf 0.912} & 0.868 & 0.785 & 0.587 \\
     & Rosetta & 14.847 & 3.100 & {\bf 0.997} & {\bf 0.983} & {\bf 0.965} & {\bf 0.930} & 0.897 & 0.831 & 0.657 \\
\hline
1H1V & F2Dock  & 29.201 & 6.849 & {\bf 0.997} & {\bf 0.984} & {\bf 0.969} & {\bf 0.937} & {\bf 0.905} & 0.844 & 0.692 \\
     & Rosetta & 38.022 & 6.123 & {\bf 0.995} & {\bf 0.972} & {\bf 0.943} & 0.884 & 0.831 & 0.727 & 0.495 \\
\hline
1Y64 & F2Dock  & 36.880 & 9.366 & {\bf 0.998} & {\bf 0.989} & {\bf 0.976} & {\bf 0.952} & {\bf 0.928} & 0.879 & 0.756 \\
     & Rosetta & 34.916 & 6.384 & {\bf 0.995} & {\bf 0.977} & {\bf 0.952} & {\bf 0.905} & 0.857 & 0.763 & 0.551 \\
\hline
3AAD & F2Dock  & 22.420 & 4.833 & {\bf 0.996} & {\bf 0.981} & {\bf 0.963} & {\bf 0.926} & 0.890 & 0.816 & 0.644 \\
     & Rosetta & 29.850 & 2.284 & {\bf 0.989} & {\bf 0.944} & 0.891 & 0.785 & 0.678 & 0.483 & 0.164 \\
  \end{tabular}
\end{table}


\section{Conclusion}
In this work, we provide a framework for providing probabilistic certificates on uncertainty in a docking algorithm. Fundamental to these certificates is the contribution of a low-discrepancy hierarchical sampling protocol that includes general amino acid information in the form of Ramachandran plots, but also structural information that is specific to the input protein in the form of a bivariate von Mises distribution. We show that the low-discrepancy samples generated by this protocol explore the energy landscape for the unbound protein, which includes samples closer to the bound conformation.

With these samples, we compare three different docking algorithms, ranging from rigid-body to completely flexible. We show that docking results vary substantially depending on the input protein structure---even for the flexible docking algorithm---further substantiating our claim that uncertainty quantification is essential to protein-protein docking.

Finally, we repeat the protein-protein docking experiments with different structures from our hierarchical sampling protocol and assess the variations in reported bound free energy. We compute a probablistic certificate for the bound free energy, and compare the 90\% confidence interval with the value computed on the bound complex. This provides a tool for comparing not only uncertainty across proteins, but also across docking algorithms.

The framework we have provided in this paper is not limited to just protein-protein docking, but can also be useful in other biological applications, where uncertainty is propagated from one stage of the algorithm to another. Notably, structure-based homology modeling uses a known, template protein structure to predict the structure of another, unknown protein with high sequence similarity. However, variations in the input structure, protein sequence alignment, and approximations in the modeling software provide many sources all lead to substantial uncertainty. If these homology-modeled proteins are then used in protein-protein docking, the uncertainty propagates and can lead to a bound free energy that has little correlation with ground truth.




\bibliography{all}

%

\end{document}